\documentclass[a4paper]{article}
\usepackage{ifthen}
\newcounter{draft}
\ifthenelse{\value{draft}=0}{
\usepackage{INTERSPEECH2016}}
{}

\usepackage{graphicx}
\usepackage{amssymb,amsmath,bm}
\usepackage{textcomp}
\usepackage{hyperref}
\usepackage{color}
\usepackage{cite}
\usepackage{time}

\DeclareGraphicsExtensions{.pdf,.png,.jpg}

\definecolor {GoogleRed}   {rgb}{0.97265625, 0.00390625, 0.00390625}
\definecolor {GoogleBlue}  {rgb}{0.0078125,  0.3984375,  0.78125}
\definecolor {GoogleYellow}{rgb}{0.9453125,  0.70703125, 0.05859375}
\definecolor {GoogleGreen} {rgb}{0.0, 0.57421875, 0.23046875}
\def\GoogleLogo{\sf \textcolor{GoogleBlue}{G}\textcolor{GoogleRed}{o}\textcolor{GoogleYellow}{o}\textcolor{GoogleBlue}{g}\textcolor{GoogleGreen}{l}\textcolor{GoogleRed}{e}}

\ifthenelse{\value{draft}=0}{
\sloppy 
\ninept

}{

}

\title{  Using instantaneous frequency and aperiodicity detection to estimate F0\\ for high-quality speech synthesis
\sthanks{The original version was accepted for presentation at 9th ISCA Workshop on Speech Synthesis, CA USA, September 13th-15th 2016.}
}

\makeatletter
\def\name#1{\gdef\@name{#1\\}}
\makeatother \name{{\em Hideki Kawahara$^{1, 2}$, 
Yannis Agiomyrgiannakis$^1$, Heiga Zen$^1$}}

\ifthenelse{\value{draft}=0}{
\address{$^1$\href{www.google.com}{\GoogleLogo} \\ 
  $^2$Wakayama University, Japan \\
  {\small \texttt{
\normalsize kawahara@sys.wakayama-u.ac.jp,\{%
    agios,%
    heigazen%
  \}@google.com %
}
}
}}
{
\author{Hideki Kawahara, Yannis Agiomyrgiannakis and Heiga Zen}
\date{\now ~BST, \today}
}

\begin{document}

\ifthenelse{\value{draft}=0}{
  \maketitle
  }{
  \maketitle
  \tableofcontents
  \listoffigures
  \clearpage
  }
\definecolor{dgreen}{rgb}{0,0.5,0}
  \begin{abstract}
  \ifthenelse{\value{draft}=0}{
}{
 {\bf \color{dgreen}[I’m coming at this as a fresh reader.  At this point I think I know what F0 (shouldn’t that be a zero?) analysis is, but I’m less sure about aperiodicity: is it, like F0, a global property of the signal at a time, or is it a “mask” on the spectrum showing which parts of the signal are not accounted for as periodic repetitions (e.g., the breathiness part)?]
 }
 {\bf \color{red} [I added short explanation of aperiodicity at the very beginning of abstract. I also returned to use F0 instead of using FO. I removed 1/5 because in this version 1/5 does not appear and replaced with ``by a factor of 10'', which is consistent with conclusion.]}
}
  This paper introduces
a general and flexible framework for F0 and aperiodicity (additive non periodic component) analysis, specifically intended for high-quality speech synthesis and modification applications.
The proposed framework consists of three subsystems: instantaneous frequency estimator and initial aperiodicity detector, F0 trajectory tracker, and F0 refinement and aperiodicity extractor.
A preliminary implementation of the proposed framework substantially outperformed (by a factor of 10 in terms of RMS F0 estimation error) existing F0 extractors in tracking ability of temporally varying F0 trajectories.
The front end aperiodicity detector consists of a complex-valued wavelet analysis filter with a highly selective temporal and spectral envelope.
This front end aperiodicity detector uses a
new measure that quantifies the deviation from periodicity.
The measure is less sensitive to slow FM and AM and closely correlates with the signal to noise ratio. 
\ifthenelse{\value{draft}=0}{
}{
 {\bf \color{dgreen} [This sounds like it’s addressing what I’ve thought of as jitter (cycle duration variation) and shimmer (cycle amplitude variation).]
 }
 {\bf \color{red} [The added note on aperiodicity solves this problem, I think.]}
}
The front end
combines instantaneous frequency information over a set of filter outputs using the measure to yield an observation probability map.
The second stage generates the initial F0 trajectory using this map and signal power information.
The final stage uses the deviation measure of each harmonic component and F0 adaptive time warping to refine the F0 estimate and aperiodicity estimation.
The proposed framework is flexible to integrate other sources of instantaneous frequency when they provide relevant information.
  \end{abstract}

\ifthenelse{\value{draft}=0}{
}{
 {\bf 
 }
}


\section{Introduction}
This paper describes a new F0 tracker for 
rapidly changing F0 trajectories with aperiodicity, which represents additive non-periodic components.
\ifthenelse{\value{draft}=0}{
}{
 {\bf \color{dgreen} [This seems like a good time to define what you mean by aperiodicity with a little more precision.  I see it’s coming in section 2, but maybe a one-sentence summary is better than assuming the word is not ambiguous.]
 }
 {\bf \color{red} [Revised the first sentence.]}
}
In high-quality speech synthesis and modification applications\cite{heiga2007details,%
kawahara2013apsipa,agiomyrgiannakis2015vocaine}, surpassing 4.2 on the 5 point MOS score, 
glitches in aperiodicity handling and the failure to follow rapidly changing fundamental frequencies (F0) are harmful to
processed speech quality. 
Introducing a generative model of F0 trajectory (for example \cite{kameoka2015generative}) to F0 estimation provides well behaved and parametric representation.
However, 
the estimated F0 trajectories are still not good enough for high-quality speech synthesis.
The actual excitation signal of speech, glottal flow, contains several sources of fluctuations\cite{titze2000principles} and consequently, the observed F0 trajectories 
are different from the trajectories 
produced by those models.
To attain highly natural synthetic speech 
it is important to retain
these fine temporal variation in F0 trajectories\cite{saitou2005development,ardaillon2015multi}.
Although many F0 extractors have been proposed\cite{boersma1993accurate,talkin1995robust,cheveigne2002jasa,kawahara2005nearly,camacho2008jasa}, 
in practice, parameter tuning and/or manual error correction is often necessary.  
In addition, their performance when extracting such fine temporal variations has not been investigated explicitly.
That is the goal of this paper.

This paper is 
organized as follows.
Section~\ref{ss:background} discusses the motivation and target for designing a new F0 observer, based on a review on existing issues.
It also defines aperiodicity, which is relevant for speech analysis and synthesis.
Section~\ref{ss:measureForObjEvaluation} presents objective measures used in this paper.
Based on these,
section~\ref{ss:architectureAndSubSystems} introduces a general scalable architecture for F0 observer.
It consists of three subsystems: front end aperiodicity detectors, the best trajectory finder, and F0 initial estimate and refinement subsystem with aperiodicity extractor.
Sub-sections~\ref{ss:estimation} and \ref{ss:refinementSubsystem} introduce the front end and the refinement subsystems, respectively.
In section~\ref{ss:evaluation},
these subsystems are evaluated using artificial test signals.
Section~\ref{ss:discussion} discusses remaining issues.
Example analysis results using actual speech samples and mathematical details are given in appendices.

\section{Background}\label{ss:background} 
Speech synthesis requires dependable F0 values whenever producing voiced sounds.
However, even for copy-synthesizing from actual speech samples, where the targets are known, this is not always easy, since voiced sounds are not purely periodic and defining F0 values for such signals is not a trivial issue.

\begin{figure}
\begin{center}
\includegraphics[width=0.9\hsize]{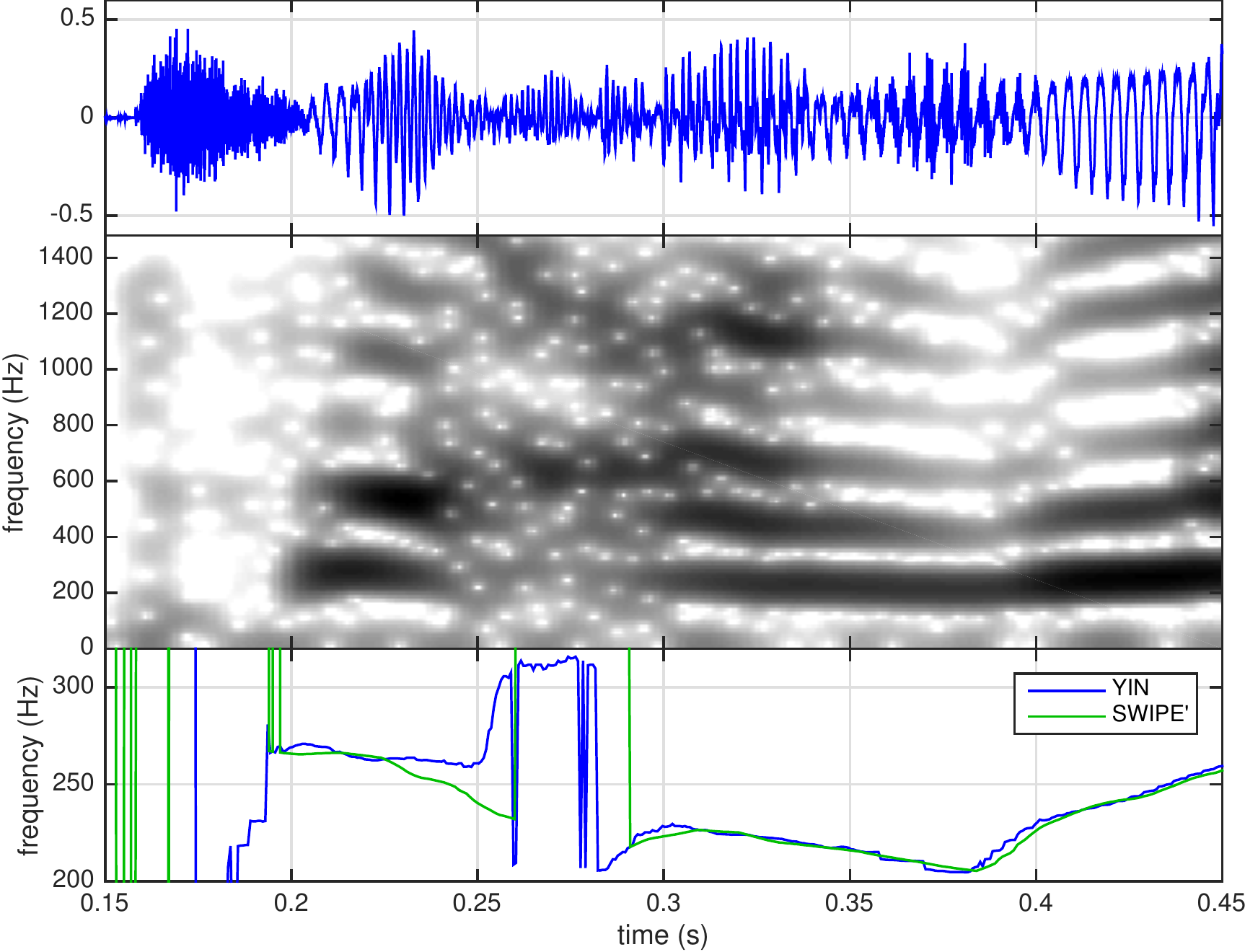} 
\end{center}
\vspace{-5mm}
\caption{Example of the difficulty of handling irregular voicing. Upper plot shows speech waveform. 
Middle shows spectrogram using 25~ms Blackman window with 1~ms frame shift.
Lower plot shows F0 trajectories extracted using YIN and SWIPE$^{\prime}$. Around 0.25~s to 0.3~s, deviations caused discrepancies and/or failure of the baseline F0 trajectory trackers.}
\label{fig:difficult} 
\ifthenelse{\value{draft}=0}{
  \vspace{-4mm}
}{
{\bf \color{dgreen} [I have no idea why this signal is so hard.  I want to see a spectrogram, although I realize that may be missing the point.  Does a zoomed-in view of the waveform give any insight into why the pitch trackers don’t agree?]}
{\bf \color{red} [Added spectrogram and stretched in time.
For ArXiV version, analysis condition of the spectrogram is added.]}
}
\end{figure}
Figure~\ref{fig:difficult} shows a beginning of a sentence from our speech corpus.
From 0.2~s to 0.52~s, the speech signal is voiced.
However, due to irregularities in glottal vibrations, defining the F0 is difficult.
The lower plot shows the F0 tracks by 
 YIN\cite{cheveigne2002jasa} and SWIPE$^{\prime}$\cite{camacho2008jasa}
 to illustrate the issues.
It is difficult to evaluate the relevance of these tracking results.
Yet these two state of the art systems do not produce consistent results.
The fact that voicing without vocal fold contact is not rare\cite{childers1986model,klatt1990vqfemaleJasa} prevents using EGG (electroglottograph) for the source of ground truth.
Using the extracted trajectory and
comparing the synthesized speech and the original speech is a reasonable test but it is very demanding on human resource and time to obtain reliable results.

An alternative approach for evaluating F0 extractors is to use an objectively defined artificial test signal. 
The ideal candidate is 
a speech signal,
where the ground truth is available and provides wide divergence and variability.
Instead, 
this article uses the excitation source signal defined by 
the L--F (Liljencrants--Fant) model\cite{Fant1985}. 
The L--F model 
represents the time derivative of the glottal flow using a set of equations with four parameters. 
However, directly digitizing the L--F model, which is defined in the continuous time domain, introduces 
spurious components due to aliasing.
To alleviate this aliasing problem
this paper uses
a closed-form representation of the anti-aliased L--F model defined in the continuous time domain\cite{kawahara2015apsipa}.
Since the model is defined in the continuous time domain, it is easy to generate a signal using a given F0 trajectory that will be the ground truth used in this paper.%
\footnote{In an open-source implementation\cite{kawahara2015SparkNG,kawahara2016interspeech} of the anti-aliased L--F model\cite{kawahara2015apsipa}, the model parameters can be controlled each glottal cycle independently to simulate the details  of vocal fold behaviour\cite{sakakibara2011physiological}.
It can be combined with a time varying lattice filter to simulate the dynamic speech production process, which modulates observed F0 through interaction between harmonic component and the group delay associated with resonances (formant trajectories).
But these detailed simulations are for further study.}
\cite{Fant1995,childers1991Jasa,garellek2014jasa}

\begin{figure}
\begin{center}
\includegraphics[width=0.9\hsize]{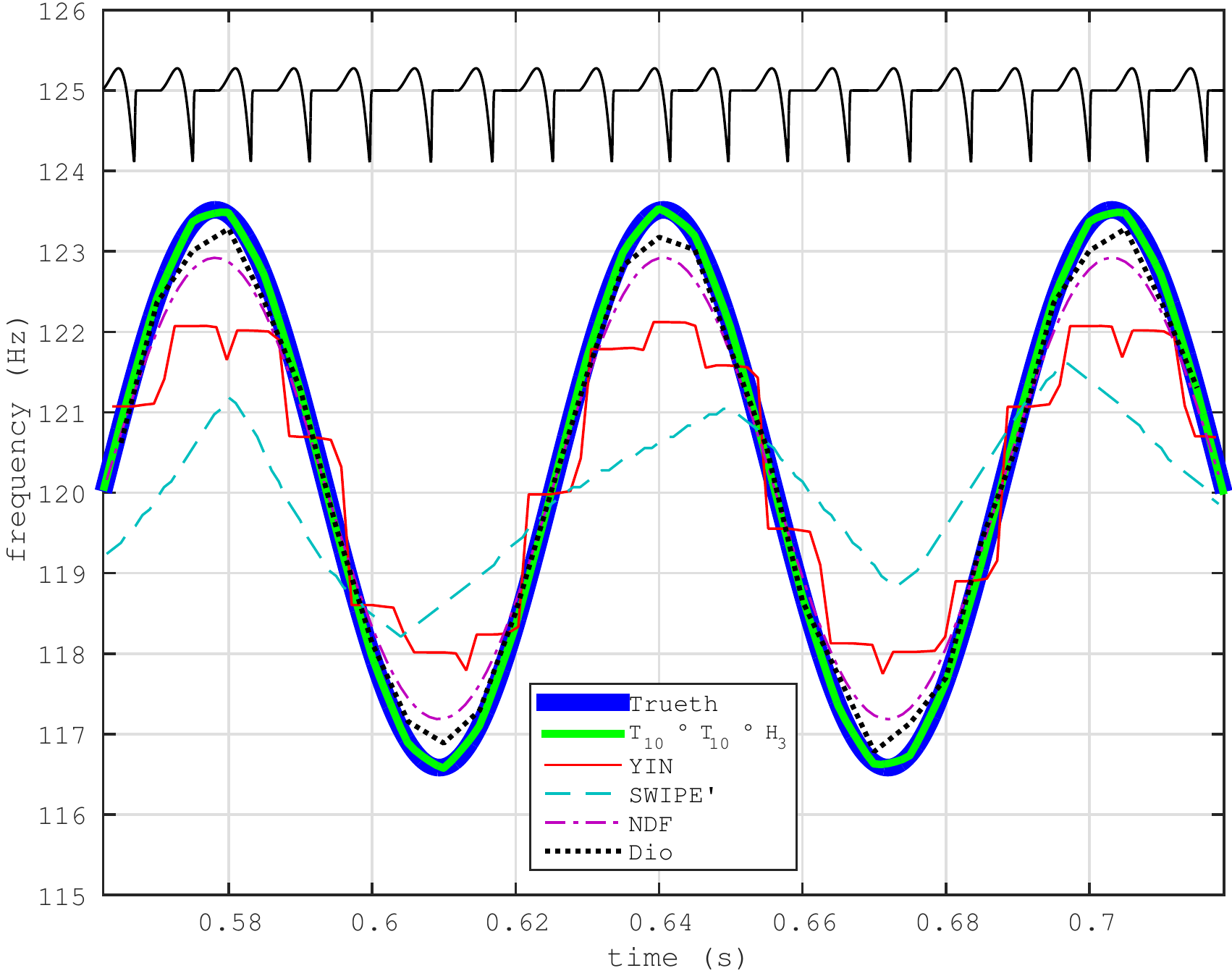} 
\end{center}
\vspace{-5mm}
\caption{Frequency modulated F0 tracking.
Black thin line on top shows waveform of the L--F (Liljencrants--Fant) model\cite{Fant1985,kawahara2015apsipa} output.
The very thick blue line shows the true F0 trajectory, which was used to generate the test signal.
The refined F0 trajectory by the proposed method (thick light green line) almost overlays on the true trajectory.}
\label{fig:trajectoryWave} 
\ifthenelse{\value{draft}=0}{
  \vspace{-4mm}
}{
{\bf \color{dgreen} [This figure does a very nice job explaining your goal and illustrating your achievement.  But it also raises alarming issues: what does it mean to vary f0 on such a rapid, sub-glottal-cycle rate?  Plus I can’t even see any variation in the period length of the waveform, so these really are subtle effects.  It’s not clear to me that the glottal flow waveform really includes sufficient information to fully recover the FM waveform, and makes me suspicious that the green line is somehow exploiting a common assumption about the form.]}
{\bf \color{red} [The waveform has enough information to recover this fine trajectory. What is needed is relevant selection of window length and Flanagan's equation.]}
}
\end{figure}
Figure~\ref{fig:trajectoryWave} shows an example of F0 tracking 
using a sinusoidally frequency modulated F0 trajectory as the test signal.
This test signal has a vibrato of 16~Hz, which is large compared to the normal human voice, but demonstrates the problems due to random, cycle-by-cycle variations in the F0.
The tested F0 extractors are YIN\cite{cheveigne2002jasa}, 
SWIPE$^{\prime}$\cite{camacho2008jasa}, NDF\cite{kawahara2005nearly}, 
DIO\cite{morise2010ieices} and the proposed method, which is described in Section~\ref{ss:architectureAndSubSystems}.
The trajectories obtained by YIN and SWIPE$^{\prime}$ are strongly distorted and attenuated, perhaps because the F0 is changing faster than these models allow.
When these distorted trajectories are used to generate the excitation source for copy-synthesis, the output is perceived differently.
This is because the distortion adds fast-changing modulation components that are not in the original signal.
The effects of these spurious components are made worse because humans are far more sensitive to fast frequency modulations than amplitude modulations\cite{tsuzaki1997jitter,bergan2001perception}.
\ifthenelse{\value{draft}=0}{
}{
{\bf \color{dgreen} [You could just put a LPF on the input to the oscillator.  The filtered version of the pitch track would still be very different from the original, but the distortion might be much less noticeable.  I know this isn’t the point, but it’s what comes to mind.]}
{\bf \color{red} [Yes, I tried LPF and reported it in Interspeech 2012 and ICASSP 2013. But, in practice, low frequency noise was damaging and bias using LPF causes made proper design of deviation measure complicated and less effective. It was dead end. I was originally inspired this by Yegnarayana's idea.]}
}

Voiced sounds are usually considered as periodic, and to first approximation the glottal pulses do occur at regular intervals.
But due to prosodic needs the F0 of a voice is constantly changing, sometimes a simple glide as in the rise of F0 in a question, and sometimes in a regular fashion, as with vibrato.
And, sometimes F0 varies in a more complex patterns, such as in tonal languages, where the F0 trajectory conveys linguistic information.
On top of these intended changes in F0, there are modulations due to physiological aspects of voice production.
The stochastic nature of neural pulses which drive the muscles of the vocal organ is a strong noise source and the critical conditions that produce vocal fold oscillation introduce bi-stable or chaotic vocal fold vibration, especially during voice onset and offset. 
Age related change and physical body status also affects the stability of vibration\cite{titze2000principles}.
All these deviations from pure periodicity play important roles in speech communication and make speech a much richer media than text\cite{fujisaki1997prosody}. 
\ifthenelse{\value{draft}=0}{
}{
 {\bf \color{blue} [Not sure what you are trying to say. Remove?
]
{\color{red} [I removed note on turbulence. It is confusing and is not very necessary here. In detailed discussions on voiced fricatives, it will be important, for example.]}
 }
}

It is important to properly analyse and replicate these deviations from periodicity in high-quality speech synthesis and modification applications.
Accurately estimating aperiodicity is still a very challenging problem.
Tracking errors introduces spurious components\cite{abe1997if,malyska2011time} and they add to the original random component.
These are the reasons why F0 tracking distortions as shown in Fig.~\ref{fig:trajectoryWave} are harmful for high-quality speech synthesis.
Two issues have to be properly solved :
accurate estimate and tracking of changing F0 trajectory and accurate estimate of random components
based on the accurate estimate of F0 trajectory.
\ifthenelse{\value{draft}=0}{
}{
 {\bf \color{dgreen} [It’s not just esimating it; at this point I’m not convinced that the problem is even well posed: How can we establish the physical reality of detailed FM trajectory in-between glottal closures?
]
{\color{red} [Glottal closure only provide F0 information in higher frequencies. The shape of airflow contributes significantly in lower frequency region and is a continuous signal.]}
 }
}

These issues motivate us to develop a framework that provides a calibrated procedure to describe the amount of aperiodicity and to track F0.                                 
The primary analysis target is high quality speech corpus recorded in a quiet and acoustically controlled environment using high-fidelity microphones.
The aim here is to provide accurate, certified metadata, in this case, F0 value and an index that represents the accuracy of the estimated F0 as well as a measure that represents the amount of aperiodicity. 
Processing speed is not the first priority of the framework 
described here.
Note
that 
these metadata
depend only on the data in the analysis frame, 
because there is no reliable model yet 
for
the dynamic behaviour of F0 and aperiodic component. 
Using 
models of dynamic F0 behaviour
such as Fujisaki's model\cite{fujisaki1998note}, or F0 continuity constraint, 
may introduce biases due to model mismatch.
Frame-based F0 with aperiodicity information, which the proposed system produces, will help to establish certifiably accurate models of the statistical/dynamic behaviour.
\ifthenelse{\value{draft}=0}{
}{
 {\bf \color{dgreen} [Not clear what this aperiodicity will measure.  I’m still not clear what we mean by aperiodicity; is the 16 Hz FM in fig 2 aperiodicity?
]
{\color{red} [No it is additive non-periodic component. The change of the first sentence of Introduction may help readers to understand this.]}
 }
}

\subsection{What is aperiodicity?}
For speech synthesis applications, amplitude and F0 are controllable 
parameters of the excitation source.
However, only replicating amplitude and F0 precisely to the original speech yields poor quality synthetic sounds.
An important attribute of excitation is missing.
This missing attribute is aperiodicity.%
\footnote{Effects of spectral envelope are also ignored.
These details exceed the scope of this paper.}
\ifthenelse{\value{draft}=0}{
}{
 {\bf \color{dgreen} [This first sentence ends up being circular, since I don’t really know what you mean by strict periodicity.  Strict periodicity is the same little cycle repeating unchanged for ever, so everything we care about at all is aperiodicity, and the term doesn’t seem that useful.

If you delete the first sentence, I think we’re fine.
]
 }
 {\bf \color{red} [Thank you. I removed the first sentence.]}
}

\ifthenelse{\value{draft}=0}{
}{
 {\bf \color{dgreen} [I added explanation about relation between OK, let me have a go, for something you can insert in the introduction or even the abstract:

Conventional model-based speech synthesis generates speech from time varying parameters describing the F0 and amplitudes in different frequency bands.  This parametric space of sounds cannot, however, achieve highly natural synthetic speech; even if we find the parameters that give the closest possible approximation of a real speech recording, there are aspects missing that cannot be captured with a local F0 and spectrum; we refer to this difference as the “aperiodic” component, and propose a model better able to reproduce them. 
]
 }
 {\bf \color{red} [Thank you. But this time, since aperiodicity is not very deeply discussed in this paper, I will use simpler current version by removing first sentence and modifying rest of the part consistently. This paper is not directly discusses model based speech synthesis. This is the second reason. I assume ``model based'' in your version refers to HMM or DNN based ones?]}
}

In this paper, attributes that can be represented by amplitude and F0 modulation are not included in the definition of aperiodicity.
What is left after removing periodic component defines ``aperiodicity'' in this paper.
It turns out that our system's F0 estimation error is well correlated with the system's estimate of aperiodicity, described below.

\ifthenelse{\value{draft}=0}{
}{
 {\bf \color{dgreen} [After all the struggle to define what aperiodicity meant, to further overload it as an index of F0 estimation performance feels wrong.  I want it to be something more like: “It turns out that our system’s F0 estimation error is well correlated with the system’s estimate of aperiodicity, described below..
]
 }
 {\bf \color{red} [Thank you. I added last sentence.]}
}

\ifthenelse{\value{draft}=0}{
}{
 {\bf \color{blue} [MS: I don't understand. HK: In speech synthesis, the source signal amplitude and frequency are temporally varying. Simply detecting deviation from pure periodicity detects deviations caused by these variation. In speech synthesis, random component is added to represents aperiodicity. If the variations caused by excitation source's AM and FM are extracted as aperiodicity and used synthesize the synthetic speech, it adds extra aperiodicity because they are already introduced by the AM and FM in the excitation source. I don't want for the definition of aperiodicity to do this double counting. I changed the last sentence of this paragraph.
 I think it is better to clearly state that aperiodicity detected in the front end consists of FM effects which will be removed in the third stage.
 The best place for this note is in Section 4.1 Front end, I think.
]
 }
}

\ifthenelse{\value{draft}=0}{
}{
 {\bf \color{blue} [No! The aperiodicity detector in Section 4.1.1 looks at any change from constant (sinusoid?) as aperiodicity. You can argue slow moving changes, that are prosodic, are attenuated.
 But your detector calls all these things aperiodicity!
]
{\color{red}
[Yes, the front end detectors detects any deviations from pure sinusoid as aperiodicity.
But at the end of the third stage,
F0 adaptive time warping makes F0 constant.
Aperiodicity detectors at this final stage are located on each harmonic frequencies.
Each detector sees only one harmonic component.
It is a sinusoid.
The detector in the final stage normalizes amplitude and the preceding time warping procedure made F0 constant.
Consequently
the detector at the very end of the stage, which is an element of $\mathcal{T}_{10}$ procedure, detects components which are not represented amplitude and F0 (on the original time axis) modulations.
This is consistent with the definition in 2.1.
]
}
 }
}

\subsection{Measures for objective evaluation}\label{ss:measureForObjEvaluation} 
F0 extractors have been evaluated based on error-rate related measures; such as Gross Pitch Error (GPE), Voicing Detection Error (VDE)\cite{chu2009reducing} and Pitch Tracking Error (PTE)\cite{lee2012interspeech}.
Attaining high performance in these measures is a prerequisite for good F0 extractors.
In this paper, we focus on F0 tracking fidelity,
because the proposed method does not make voiced/unvoiced decision.
Instead, 
this F0 tracker
outputs a measure of aperiodicity, which closely correlates with the standard deviation of the relative F0 estimation error from the true value.
This aperiodicity detector 
also is an informative source of the type of excitation.
The voiced/unvoiced decision is left to the application, which can use the output of the proposed method to make this decision.

\section{Architecture and subsystems}\label{ss:architectureAndSubSystems} 
The proposed framework, YANGSAF (Yet ANother Glottal Source Analysis Framework), computes the instantaneous F0 using three steps: estimate, track, and refine.  The estimation step calculates three features of the input signal over a number of bandpass channels. The maximum from the estimate stage is then tracked to produce a local estimate of the F0. Finally, an optional refinement stage combines temporal and harmonic information to produce a more accurate estimate of F0.

\subsection{Estimation}\label{ss:estimation} 
\begin{figure}
\begin{center}
\includegraphics[width=0.9\hsize]{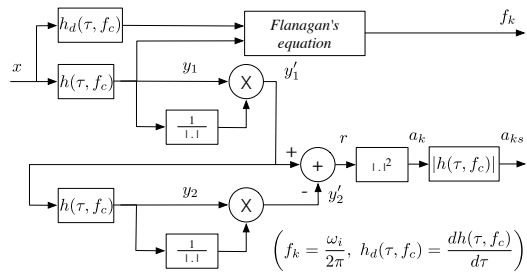} 
\end{center}
\vspace{-6mm}
\caption{Schematic diagram of aperiodicity detector.
Upper part calculates instantaneous frequency using Flanagan's equation (Appendix~\ref{ss:flanagansEq}).
The lower part calculates aperiodicity measure as a relative residual level $a_{ks}$ (Appendix~\ref{ss:residualForFrontend}).}
\label{fig:detector} 
\ifthenelse{\value{draft}=0}{
  \vspace{-4mm}
}{
}
\end{figure}

The first stage of the YANGSAF algorithm analyses the signal with a number of bandpass channels, and then estimates three values for each channel as a function of time. These values are 1) the local instantaneous frequency, 2) a measure called aperiodicity that represents the amount of variability in the channel's frequency estimate, and 3) a probabilistic estimate that the channel contains a good representation of the F0. These signals are described in the subsections that follow and are used in the tracking stage described by Section 4.2. Figure~\ref{fig:detector} shows a diagram of the estimating detector in each channel.

The front end breaks the input into a number of spectral channels using a bank of bandpass filters, each centered at $f_c$.\footnote{The -3~dB points in frequency are $0.745f_c$ and $1.255f_c$. The zero points are located at 0 and $2f_c$. The -3~dB points in time are $-0.456/f_c$ and $0.456 f_c$. Support is $(-2 /f_c, 2/f_c)$.
} 
The center frequencies cover the possible F0 range, with a fixed separation on the logarithmic frequency axis. The current implementation covers 400~Hz to 1000~Hz using 12 channels and detectors in each octave. 
\ifthenelse{\value{draft}=0}{
}{{\bf \color{dgreen} [Why -40dB points?  I’m more interested in -3 dB points.  Actually what I’m interested in is the impulse response (i.e how much blurring in time it imparts), since we’re basically “doing Fourier’s work” by converting our nice glottal impulse train into a complex sinusoid by laying down a little chunk of sinusoid at each glottal pulse.]}
{\bf \color{red}
 [I changed to -3~dB points and showed both frequency and time width.]}
}

The instantaneous frequency estimate needs both the complex-valued signal and its derivative. 
These values are calculated starting with bandpass filter $h(\tau, f_c)$ and its derivative $h_d(\tau, f_c)$ shown in Fig.~\ref{fig:detector} and described in Appendix~\ref{ss:flanagansEq}. 
Each bandpass filter has linear phase, is a zero-delay FIR filter, has a complex-valued response, and passes only the positive frequency components. 

\begin{figure}
\begin{center}
\includegraphics[width=0.85\hsize]{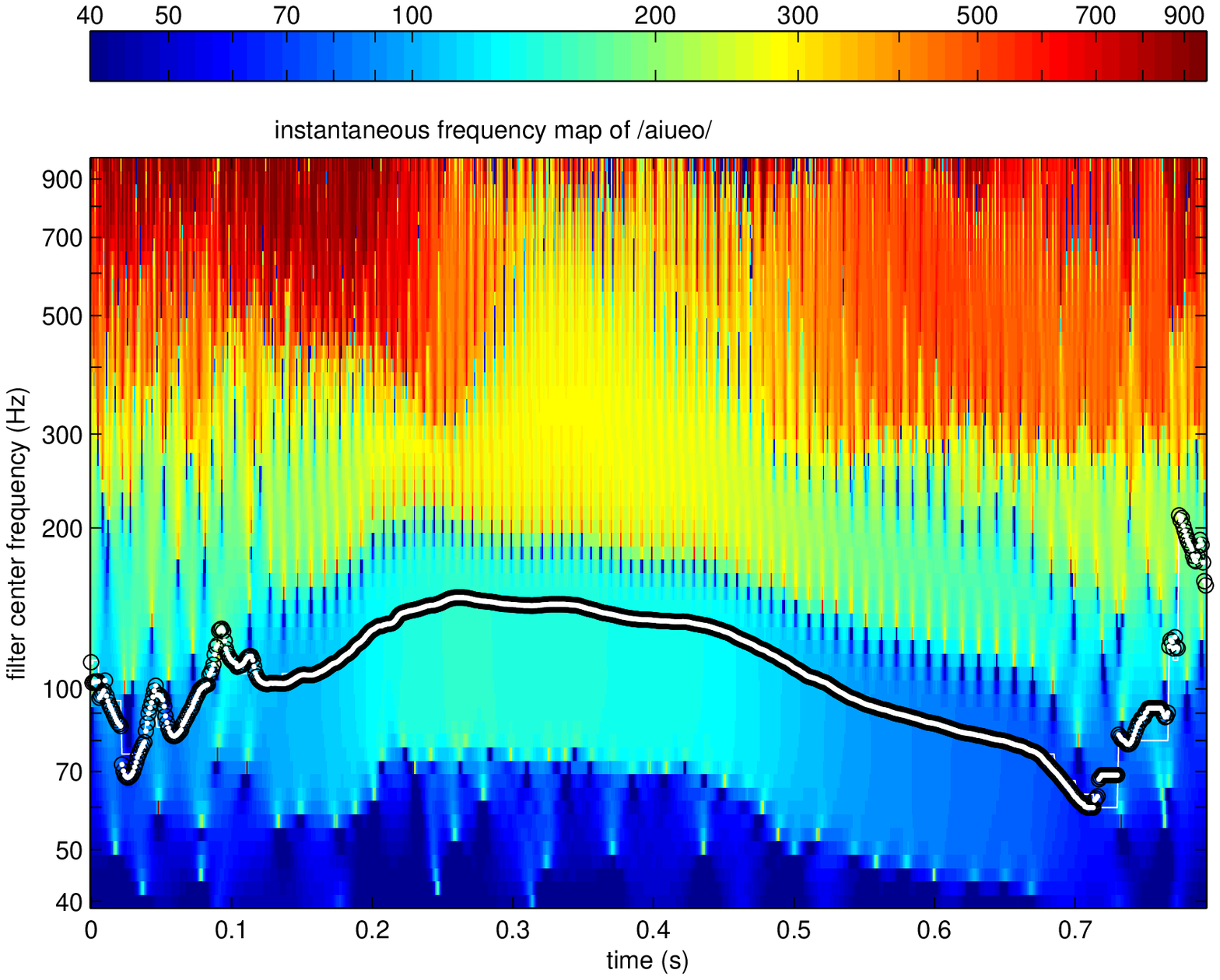} 
\includegraphics[width=0.85\hsize]{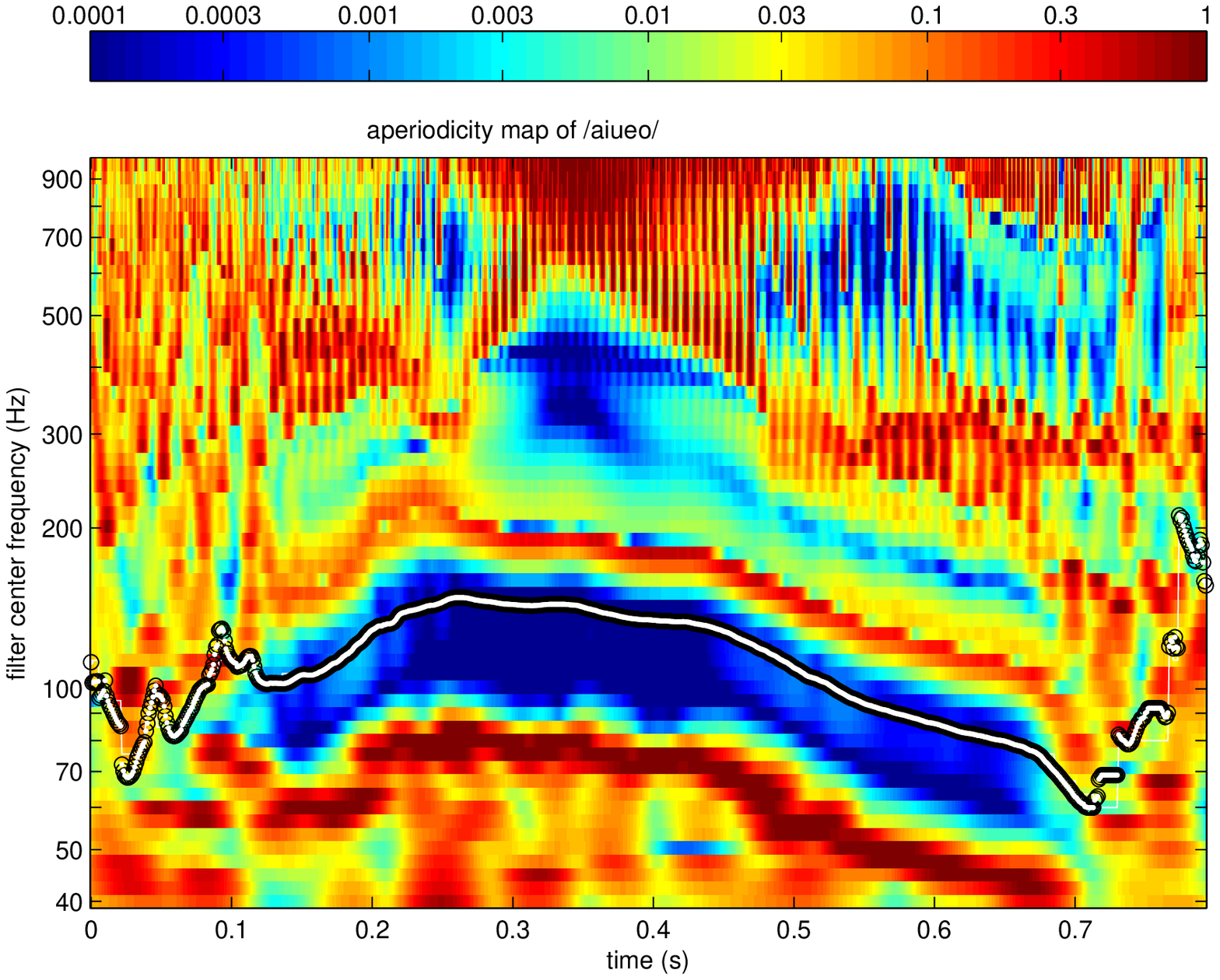} 
\includegraphics[width=0.85\hsize]{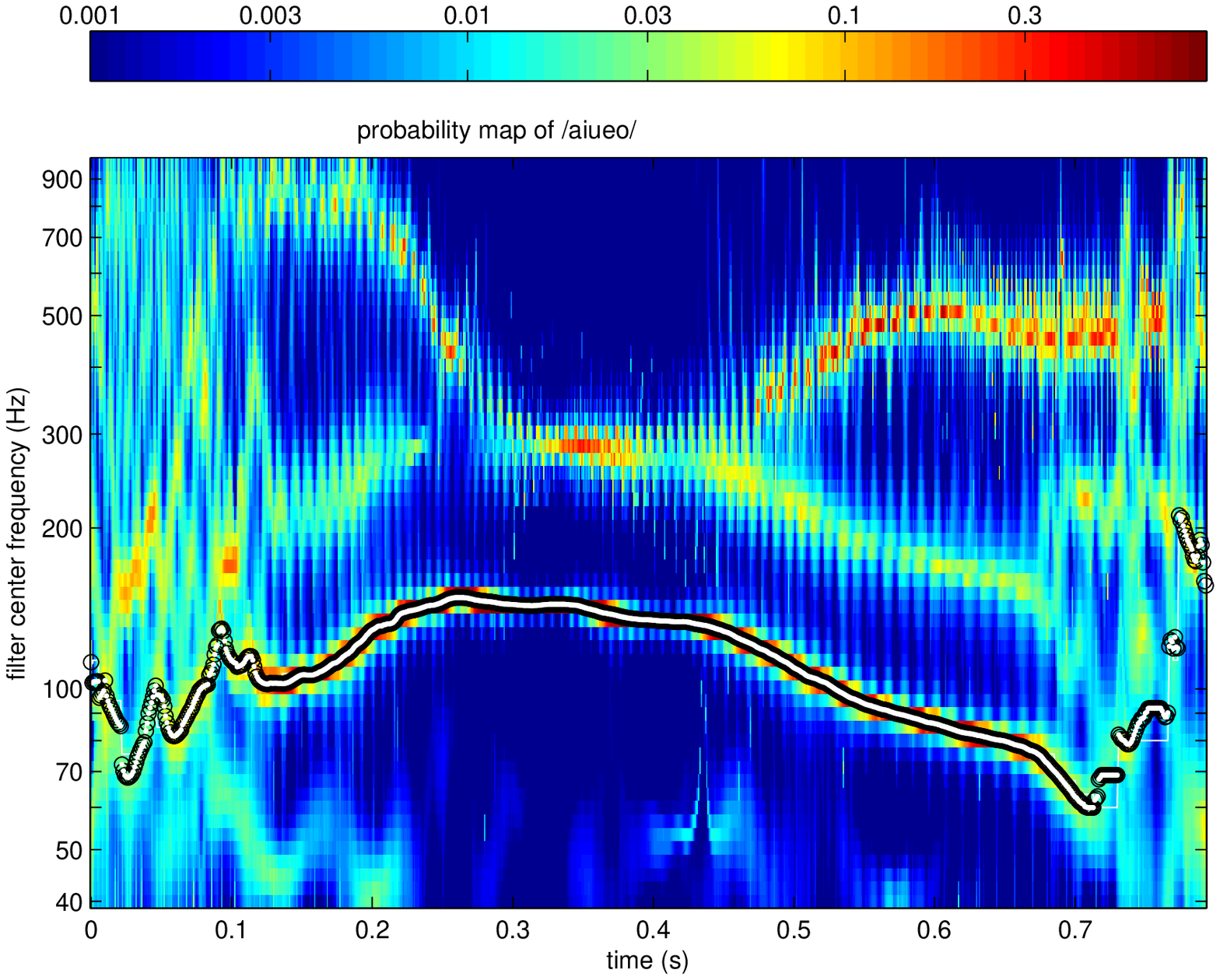} 
\end{center}
\vspace{-5mm}
\caption{Example of the first stage detector outputs. 
The upper plot shows the instantaneous frequency map.
The middle plot shows the residual map.
The bottom plot shows the probability map.
The speech material is a Japanese vowel sequence /aiueo/ spoken by a male.
For reference purpose, the F0 trajectory extracted in the third stage is overlaid using open circles.
 In the probability map, the periodic vertical lines are synchronized with vocal fold vibration. 
 The upper right trace of periodicity corresponds to the response of first formant of vowel /o/. 
}
\label{fig:vaiueoEx} 
\ifthenelse{\value{draft}=0}{
  \vspace{-3mm}
}{
}
\end{figure}
Figure~\ref{fig:vaiueoEx} shows an example of these three estimated signals for a sequence of vowels.

\subsubsection{ Instantaneous Frequency}

The instantaneous frequency of the signal contained within each channel is calculated using Flanagsn's approach, which is based on the logarithm of a complex signal $x(t)$ and its derivative. An AM/FM modulated signal is represented in polar form $x(t) = r(t)e^{j\theta(t)}$.
The instantaneous (angular) frequency $\omega_i(t)$ is defined as the derivative of the phase component $\theta(t)$, namely $\omega_i(t) = \frac{d\theta(t)}{dt}$.
The instantaneous frequency can be derived by starting with the logarithm of the component phase and using a bit of algebra: \vspace{-3mm}
\begin{align}
\!\!\! \dfrac{d\log(x(t))}{dt} & = \! \dfrac{d\log\left(r(t)e^{j\theta(t)}\right)}{dt} \!
= \! \dfrac{d\log(r(t))}{dt}\! + \! j\dfrac{d\theta(t)}{dt} \\
\omega_i(t) 
& = \frac{\Re[x(t)]\dfrac{d\Im[x(t)]}{dt}-\Im[x(t)]\dfrac{d\Re[x(t)]}{dt}}{|x(t)|^2}\label{eq:flanagan} ,
\end{align}
where $\Re[x]$ and $\Im[x]$ represents the real and the imaginary part of $x$, respectively. The derivation of this expression is contained in Appendix~\ref{ss:flanagansEq}.

\subsubsection{Aperiodicity}

We also wish to calculate a measure of the aperiodicity of the signal in each channel, which will be used as a measure of the reliability of the instantaneous frequency measurement. 
For a constant sinusoid, the aperiodicity is zero, and the aperiodicity grows as the signal varies (wiggles) more within the bandpass channel. The basic idea of the periodicity detector is to calculate the amount of energy in the band-passed signal that is {\em not} the primary sinusoid. The primary sinusoidal component will have the largest energy, and when the complex signal is normalized to have unit magnitude, refiltered, and then renormalized, the primary sinusoid will still have unit magnitude. The other components will be filtered with a non-unit gain, since the filter is not an ideal brick-wall filter, and their amplitude will change. Subtracting the original and the twice-filtered and normalized response gives an estimate of the aperiodicity. Note this estimate is done {\em without} explicitly identifying the primary sinusoid and its frequency.

When a signal $x$ whose fundamental frequency is equal to $f_c$ is filtered, only the fundamental component, a complex-valued, slowly time-varying signal, is passed (appears in $y_1$) and is normalized to become $y_1^{\prime}$. 
Then, by using the same filter, filtering signal $y_1^{\prime}$ again, and normalizing the overall amplitude using the absolute value of the complex valued-signal, the twice filtered (and amplitude normalized) signal $y_2^{\prime}$ is obtained. Subtracting this twice filtered and amplitude normalized signal $y_2^{\prime}$ from the amplitude normalized first filter output $y_1^{\prime}$ , yields a residual signal $r$. 
Since the signal $y_1^{\prime}$ is normalized, the power of the residual represents the relative level of the other component(s). 
\ifthenelse{\value{draft}=0}{
}{{\bf \color{red}
 [I replaced ``random'' in the last sentence with ``other''.]}
}

The difference between $y_1^{\prime}$ and $y_2^{\prime}$ corresponds to spectral components in the channel that are not the primary sinusoid. Calculating the energy in this signal $(a_k)$, and smoothing it gives $a_{ks}$ which is this system's measure of harmonic aperiodicity. Appendix~\ref{ss:residualForFrontend} describes the relation between the SNR of the original signal and the residual aperiodicity power using equations and examples. 
\ifthenelse{\value{draft}=0}{
}{{\bf \color{blue} [A little redundant. Can fix later when we know page budget.]

[Haven’t said anything about how the lowest frequency channel is likely to have the most energy, and thus you’ll choose the fundamental, not the higher harmonics, right????]

{\color{red} [I added the following paragraph to explain it.
But you explained this at the beginning of probability section. 
I think the following paragraph is redundant and may be removed.]}
}
}

Placing bandpass filters having the same shape on the logarithmic frequency axis yields the detector to output higher aperiodicity value, when $f_c$ is located at harmonic frequencies other than the fundamental.
This is similar to the concept ``fundamentalness,'' which is explained in Fig.~11 of reference~\cite{kawahara1999spcom}.
Appendix shows relation between filter shape examples and harmonic components.

The instantaneous frequency calculation and the aperiodicity calculation yield values at the audio sampling rate. These audio sampling rate time series are down-sampled for later processing. In this work the down-sampling is accomplished by extracting the nearest time samples from each time series, providing two sequences of instantaneous frequency and aperiodicity measure values at the frame rate (i.e. 200 Hz).

\subsubsection{Probability}\label{ss:ssprobability} 
The fundamental component in the original signal is dominant in a number of output channels because there is little else for filters centered at frequencies lower than the second harmonic can respond. Thus a number of channels will respond in the same way to the fundamental component, as seen by the blueish blob around 100Hz in the second panel of Figure~\ref{fig:vaiueoEx}. All channels inside this blob have information about the fundamental component, but with different reliabilities. 

Given a number of (distinct) estimates of the true F0, all from different channels, a probability map indicates which channel will have the best estimate. To create this probability map, all the instantaneous frequency and aperiodicity estimates are converted into Gaussian probability masses centered at various instantaneous frequency estimates. The output of the channel's aperiodicity estimate ($a_{ks}$, a measure of smoothed energy) is converted into a variance $\sigma_k^2$ by scaling.
The scaling coefficient was empirically determined by a set of simulations.
On a log-frequency scale $\nu$, this gives a number of (independent) estimates of the instantaneous frequency, each modelled as a Gaussian mass centered at $\log(f_k)$, and with a variance of $\sigma_k^2$. 
Summing all these yields a probability density function $p_{G}(\nu)$ represented as a Gaussian mixture.
For each channel, 
integrating this distribution provides
an observation probability $P_r[k]$ that channel $k$ should see the fundamental component in its nominal pass band $[f_L(k), f_H(k)]$ is
\begin{align}\label{eq:totalProb} 
p_{G}(\nu) & =  \sum_{n=1}^N \!\frac{\hat{b}_n}{\sqrt{2\pi \sigma^2_n}}  \exp\!\!\left(\! - \frac{\left(\log(f_n)\! -\! \nu\right)^2}{\sigma_n^2}\! \right) \! \\
\!\!\!P_r[k] & = \!\! \int_{\log(f_L[k])}^{\log(f_H[k])}\!\! p_{G}(\nu) d\nu \\
\!\!\!f_L[k] & = f_c[k]2^{-\frac{1}{2K}} , \ \ f_H[k]  = f_c[k]2^{\frac{1}{2K}} ,
\end{align}
where $K$ represents the number of filters per octave.
\ifthenelse{\value{draft}=0}{
}{{\bf \color{dgreen} [Can't you just plot the mixture of Gaussians?  Why do you have to integrated it?  Is it because some of the Gaussians degenerate into deltas, so we want to (effectively) bin them into semitone bins?  You could get a similar effect by disallowing $\sigma_n$ to become smaller than ~ 1 semitone bin.]

{\color{red} [Yes I can and it does. It is also a practical decision, since channel based trajectory is used later.

For ArXiV version, I explicitly inserted a Gaussian mixture pdf.]}
}
}
   
This integrates the instantaneous frequency probability distributions between the frequency limits of filter $k$ to arrive at an estimate of how reasonable it is for channel $k$ to provide an estimate of the F0. An example of this result is shown in the bottom of Figure\ref{fig:vaiueoEx}.

\subsection{Tracking}\label{ss:bestTrajectory} 
Given the three instantaneous maps (as a function of frame time and spectral channel) computed in Section~\ref{ss:estimation}, an initial estimate of the single best F0 at each frame is calculated by finding the channel with the highest probability. This is done in four steps: estimate the pitch range for this utterance, smooth the probability map, find the highest probability F0, and then refine the F0 estimate. The result is a smooth estimate of the true F0 based on the instantaneous frequency calculated in each channel.

First, the F0 search range is estimated by a weighted average of the instantaneous frequencies seen in the utterance. The temporal weighting is calculated from the energy in the original signal, after filtering it between 40-1000Hz, which is the prospective pitch range. Then each frame of the instantaneous frequency map is weighted and combined to form an overall instantaneous frequency histogram. By weighting by the signal's amplitude at each point in time, the high-energy portion of the utterance (vowels) are treated with more importance.

The median of this instantaneous frequency distribution (marginal distribution) defines the center point of the F0 search range. The tracker looks for peaks in the probability distribution within 1.2 octaves above this center point, and 1.3 octaves below, a total of a 2.5 octave range.
\ifthenelse{\value{draft}=0}{
}{{\bf \color{dgreen} [This seems quite hacky; how much does it improve results compared to a fixed, wide search range?]

{\color{red} [This is a fixed range. I have to make clear of it. I added ``marginal distribution.'']}
}
}

Second, in order to better estimate the F0 at the start and end of voicing the probability map computed in Section~\ref{ss:ssprobability} is smoothed
in time using a 45ms Hanning window with amplitude weighting.
Smoothing is done before tracking
so that we extend the F0 estimates at the start and end of voiced segments.
For example, at the onset of voicing, the probability at F0 is not high, because the signal level is low and the SNR is low. 
Smoothing using amplitude weighting increases the probability at F0, because at frames after the onset the level grows and consequently the SNR become higher. In other words, the probability distribution of the onset frames become more like the probability distribution of later frames. 
This way smoothing reduces tracking error at the beginning of voicing. The same thing happens at the voice offset. 

Thirdly, given the F0 range and the smoothed probability map the best channel across time can be tracked.
For a range of channels that are within the 2.5 octave range defined for the entire utterance, and 0.7 octaves of the last frame’s best channel, the channel with the highest smoothed probability is chosen. 

Finally, this channel selection is further refined by returning to the original probability map computed in Section~\ref{ss:ssprobability} and choosing the channel with the highest probability closest to that bin chosen from the smoothed estimate.
The following provides the initial F0 estimate $f_{OI}$.
\begin{align}
f_{OI} & = \sum_{m \in \mathbb{V}[k]} b_m f_m  \\
\mathbb{V}[k] & = \{m \ |\ 0.5\ f_c[k] < f_c[m] < 1.25\ f_c[k] \ \} \ \ ,
\end{align}
where the best weights $b_m$ are calculated from $\sigma_m^2$ in $\mathbb{V}[k]$.
\ifthenelse{\value{draft}=0}{
}{
 {\bf \color{dgreen}[To do this kind of continuous, time-varying resampling when the objective is to result in very low-distortion results seems challenging.  But maybe that’s covered in the refs?]
 }
 {\bf \color{red} [Careful investigation is the next step for precise aperiodicity analysis. Upsampling using Nuttall window and then linear interpolation for F0 adaptive time stretching will work. But for F0 refinement only, simple linear interpolation worked.]}
}

\subsection{Refinement of the initial estimate}\label{ss:refinementSubsystem} 
The third stage further improves this F0 estimate by adding two refinements.
First, and most importantly, the higher harmonics of an F0 estimate can refine the estimate.
Secondly, adaptive time warping of the original signal, combined with further refinement using higher harmonics of the warped signal, reduces the amount of F0 trajectory deviation for better analyses.
\ifthenelse{\value{draft}=0}{
}{
 {\bf \color{red}[I added ``combined with further refinement using higher harmonics of the warped signal'' to make clear.]
 }
} 

The first procedure uses harmonic frequencies and their variance.
Each harmonic component, from first to $m$-th, has corresponding aperiodicity detector.
Each bandpass filter of the detector has the same shape on the linear frequency axis and does not cover neighbouring harmonic components.
Each detector yields instantaneous frequency $f_k$ and its aperiodicity $a_k$, where $k$ represents the harmonic number.
These values are converted to F0 estimate $f_k/k$ and its variance $\sigma^2_k$.
The weighted average $\sum_{k=1}^m b_k f_k/k$ provides the refined F0 estimate.
Variance values $\{\sigma^2_k \}_{k=1}^m$ are used to calculate the best mixing weights $\{b_k \}_{k=1}^m$ (Appendix~\ref{ss:bestMix}).
\ifthenelse{\value{draft}=0}{
}{
 {\bf \color{red}[I rewrote this paragraph.]
 }
} 

However, this refinement does not properly make use of higher harmonic information when the F0 trajectory is rapidly changing.
This is because a rapid movement of higher frequencies generates strong side-band components and they smear the analysed harmonic structure\cite{abe1997if,Kawahara99fixedpoint,malyska2011time}.

Thus, the second procedure uses F0 adaptive time axis warping to alleviate this problem.
Stretching the time axis, proportional to an instantaneous F0 value makes the observed F0 value 
 constant\cite{abe1997if,Kawahara99fixedpoint,malyska2011time} and
keeps the 
harmonic structure intact.
Then, placing aperiodicity detectors on harmonic frequencies, from first to $m$-th, the weighted average of F0 information yields the F0 estimate on the warped time axis.
Converting this estimate value to the value on the original time axis provides the further improved F0 estimate.
\ifthenelse{\value{draft}=0}{
}{
{\bf \color{blue} [The description above doesn't mention anything about harmonics when time stretching.?]}
 {\bf \color{red}[I rewrote latter part of this paragraph by adding notes on the use of harmonic components in the warped time axis.]
 }
}

These two procedures are applied serially as well as recursively.
Let $\mathcal{H}_m$ represent the operation of harmonic based refinement using the first through $m$-th harmonic components and $\mathcal{T}_m$ represent the operation of F0 adaptive time warping-based refinement using the first through $m$-th harmonic components.
Let $\mathcal{P}_X[x; \Theta ]$ represent the function of initial estimate F0 where $x$ represents the input signal and $\Theta$ represents a set of the associated design parameters for analysis.
The following equations describes the configurations of the two trackers tested:
\begin{align}
\mathcal{H}_{10} \circ \mathcal{H}_{3}  \circ \mathcal{P}_X[x ; \Theta ] &
\label{eq:harmonicBasedRef}  \\
\mathcal{T}_{10} \circ \mathcal{T}_{10} \circ \mathcal{H}_{3}  \circ \mathcal{P}_X[x ; \Theta ] &
\label{eq:refWithTwTw}  ,
\end{align}
where $\mathcal{T} \circ \mathcal{H}$ represents the composite function of the functions $\mathcal{T}$ and $\mathcal{H}$.

Finally, by placing aperiodicity detectors on all harmonic frequencies in the warped time axis, estimated SNR around each harmonic component provides the excitation source information for speech synthesis.
Because any F0 trajectories on this warped time axis are constant in time, aperiodicity values which detectors output are consistent with the aperiodicity definition of this paper.

\section{Evaluation using test signals}\label{ss:evaluation} 
This paper uses two measures of performance.
Most importantly, the standard deviation of the relative error tells us the total distortion of the estimated F0 trajectory from the ground truth.
The second performance measure is the frequency-modulation amplitude transfer function (FMTF), 
which expresses how well a F0 tracker follows fast F0 modulations.
The test signal uses sinusoidal modulation on the logarithmic frequency axis,
since F0 dynamics is better described on the logarithmic frequency axis\cite{fujisaki1998note}.
Consequently, both FMTF and distortion evaluation measures use logarithmic frequency to calculate their value.

The proposed algorithms are implemented using \textsc{Matlab} and tested using synthetic signals.
Only representative results are described below. 
In the following tests, the test signals were generated using the aliasing-free L--F model\cite{kawahara2015apsipa}.%
\footnote{The original L-F model\cite{Fant1985} is anti-aliased using a closed form representation. The \textsc{Matlab} implementation of this function and GUI-based interactive application for speech science education are open source\cite{kawahara2015SparkNG,kawahara2016interspeech}.
Spurious levels around the fundamental component of the model's output are lower than -120~dB.}
The sampling frequency $f_s$ was 22050~Hz 
and the ``modal'' voice quality parameters\cite{childers1995jasa} for the L--F model were used in the following examples.
\ifthenelse{\value{draft}=0}{
}{
 {\bf \color{dgreen}[This evaluation seems to depend critically on how this model handles F0 variations between GPEs.  Since you’re using it, I assume it does something consistent with your underlying hypothesis of a continuously-varying F0 with meaningful values between pitch pulses.  But as a reader, not knowing what it does is a handicap.]
 }
 {\bf \color{red} [Thank you for pointing out this. I will add appendix for the arXiV version.]}
} 

We test this new F0 tracker in two different ways: additive noise and FM modulation.

\subsection{Additive noise}
Firstly, the quality of the F0 estimate in the face of additive white noise 
was tested using the configuration given by Eq.~\ref{eq:harmonicBasedRef}
 ($\mathcal{H}_{10} \circ \mathcal{H}_{3}$).
The F0 extractor for the initial estimate (Section~\ref{ss:bestTrajectory}) ($\mathcal{P}_X[x ; \Theta ]$) was tested to clarify the effects of refinement (Section~\ref{ss:refinementSubsystem}).
Four popular F0 extractors were also evaluated for reference;
YIN\cite{cheveigne2002jasa}, SWIPE$^{\prime}$\cite{camacho2008jasa}, NDF\cite{kawahara2005nearly} and DIO\cite{morise2009fast,morise2010ieices}.
They were tested using their default or recommended settings.
A constant F0 trajectory was used in this test.

\begin{figure}
\begin{center}
\includegraphics[width=0.9\hsize]{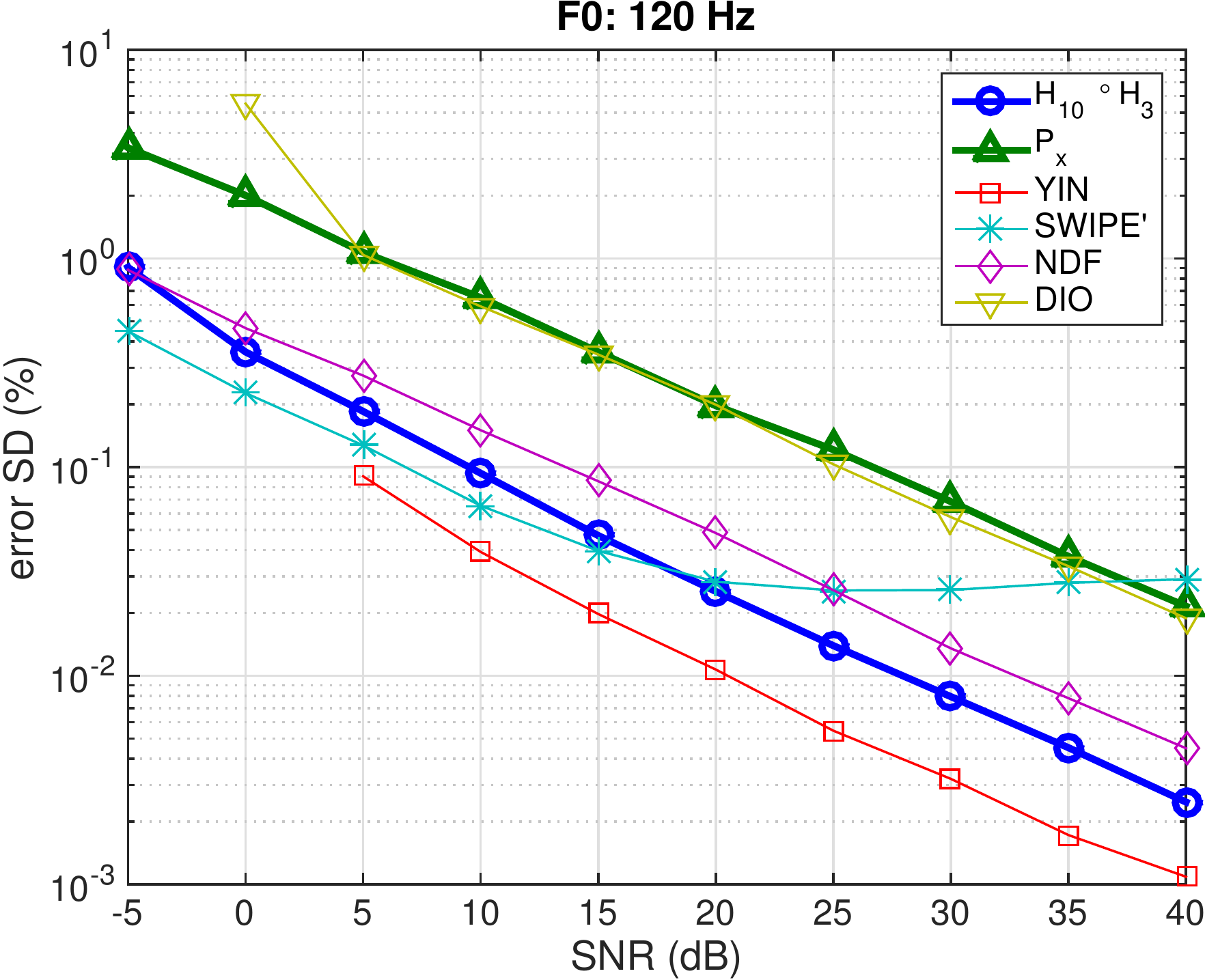} 
\end{center}
\vspace{-5mm}
\caption{RMS error of F0 estimation vs. additive noise SNR for a temporally constant F0.
The initial estimate (triangle) error deviations were reduced by a factor of 8 (circle) by using harmonic refinement.}
\label{fig:snrAddG} 
\ifthenelse{\value{draft}=0}{
  \vspace{-3mm}
}{
}
\end{figure}
Figure~\ref{fig:snrAddG} shows the results for a 120~Hz F0.
The vertical axis represents the relative RMS error.
When the SNR is larger than 5~dB, YIN yielded the best results.
But, YIN's performance is obtained at the cost of poor temporal resolution, which will be shown in the following test.
DIO was designed for high-quality recordings and is not tolerant to noise. 
While SWIPE$^{\prime}$ showed good performance from 0 to 20~dB SNR, performance saturated there after.
The harmonic refinement procedure reduced the error in the initial estimate by a factor of 8, even in high noise, because the standard deviation of error in $n$-th harmonic component is $1/n$ as described in previous paragraph.
In total, this is the second best result.
\ifthenelse{\value{draft}=0}{
}{
 {\bf \color{dgreen}[Not sure what you mean here.  You mean that by combining estimates from N harmonics, each with random noise, we can reduce the noise variance by 1/N?.]
 }
 {\bf \color{red} [The estimated instantaneous frequency of the $n$-th component is $nf_0$. To get information about F0, it is necessary to divide the estimated instantaneous frequency by $n$. The standard deviation also has to be multiplied by $1/n$. But, I am still uneasy about this. Something is wrong.]}
}

\subsection{Frequency modulation of F0}
Measuring the ability of a F0 tracker to follow F0 modulation is a more relevant test for speech sounds with rapid changes.
The instantaneous frequency of the aliasing-free L--F model output was controlled at audio sampling rate (22050~Hz) resolution.
The average F0 was 120~Hz with 100~musical cent peak-to-peak modulation depth roughly to 6\% frequency modulation peak-to-peak in frequency.%
\footnote{Tested F0 were 120, 240, 480 and 800~Hz.
For F0 extractors, 120~Hz is the worst condition in terms of tracking.}
In  the two tests described in this section, a bit of white noise (SNR 100~dB) was added.
\ifthenelse{\value{draft}=0}{
}{
 {\bf \color{dgreen}[The reader wonders: Why add noise?.]
 }
 {\bf \color{red} [SWIPE did not work in some conditions. I was not able to find the reason. But adding small noise solved this anomaly. This is a hack which I cannot explain the reason and simply wrote what I did.]}
} 

\begin{figure}
\begin{center}
\includegraphics[width=0.9\hsize]{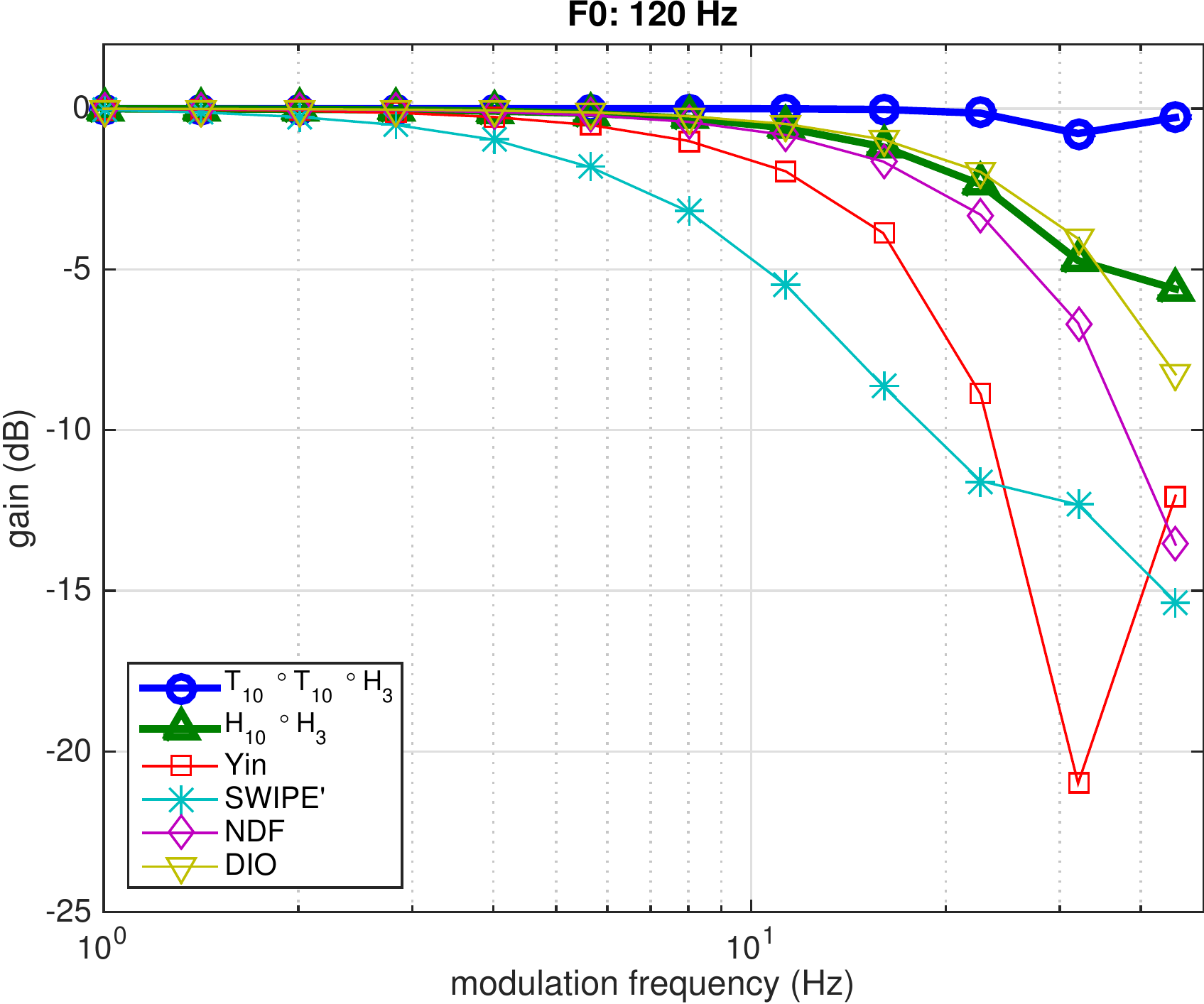} 
\end{center}
\vspace{-5mm}
\caption{Frequency modulation transfer function for F0 modulation. 
The higher tracking frequency limit of the initial F0 estimate (triangle) is expanded two times by the proposed refinement using F0 adaptive time warping (circle).}
\label{fig:modulationTF} 
\ifthenelse{\value{draft}=0}{
  \vspace{-3mm}
}{
}
\end{figure}
Figure~\ref{fig:modulationTF} shows the frequency modulation transfer function for the four F0 trackers that serve as a benchmark and two variations of the F0 tracker described in this paper.
For very low vibrato frequency (low modulation frequency) all F0 trackers work well at high SNR.
At higher modulation frequencies all F0 trackers except for $T_{10} \circ T_{10} \circ H_3$ fail to follow the full modulation, which shows up as a reduced gain when considering the output vs input modulation deviation.
For higher F0 signals, the 3~dB point increased proportionally to the F0 value, except YIN.

\begin{figure}
\begin{center}
\includegraphics[width=0.9\hsize]{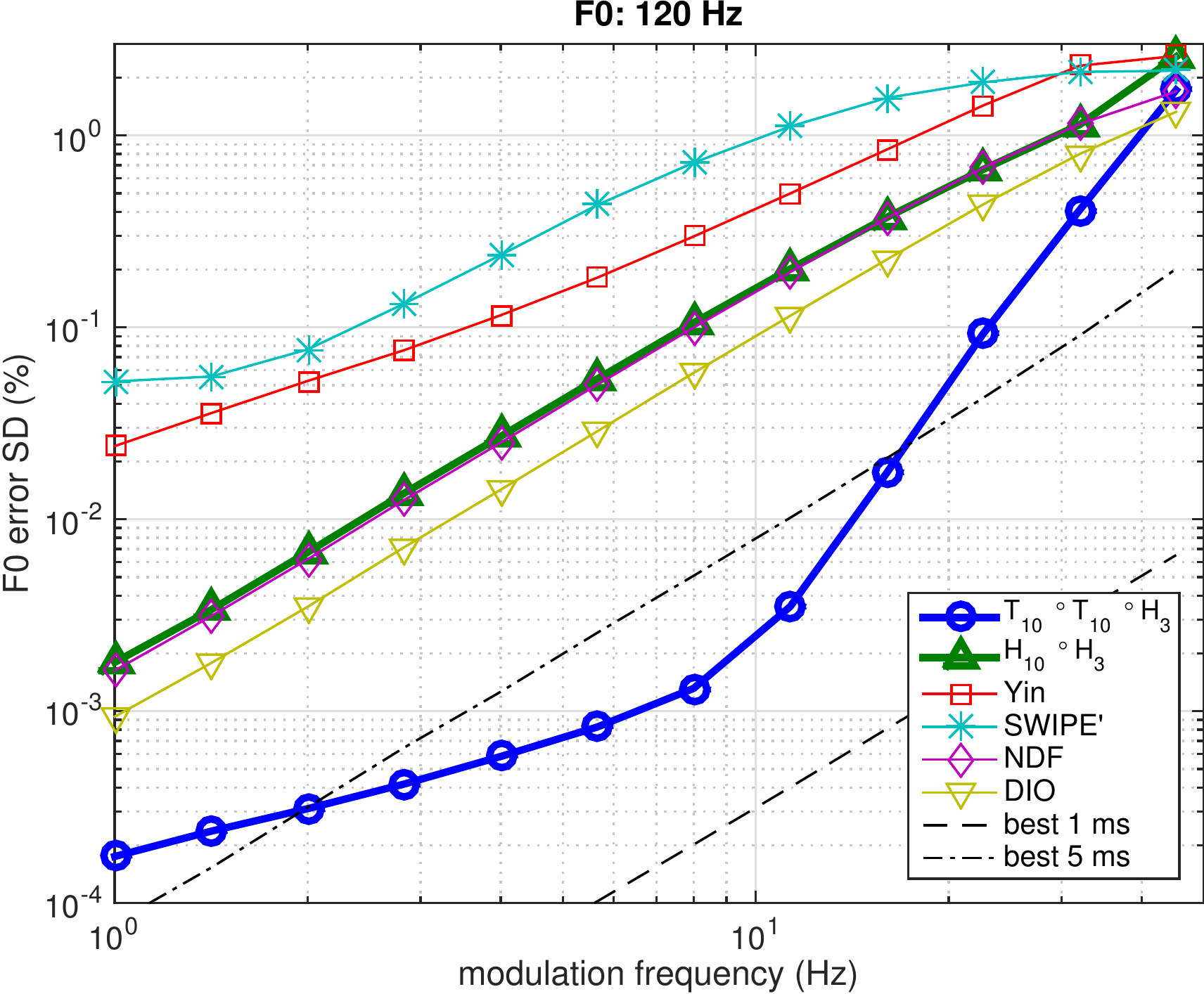} 
\end{center}
\vspace{-5mm}
\caption{RMS error of F0 trajectory tracking.
The RMS error of the refined F0 trajectory using harmonic frequencies
(triangle) is reduced by a factor of 10 or more by introducing F0 adaptive time warping (circle).
}
\label{fig:errorSD} 
\ifthenelse{\value{draft}=0}{
  \vspace{-4mm}
}{
}
\end{figure}
Figure~\ref{fig:errorSD} shows the RMS error of the F0 trajectories as a function of the modulation frequency.
The dashed line and dash dot line show the RMS error of the best approximation to the true F0 using piece-wise linear function with segment lengths 1~ms and 5~ms respectively.

SWIPE$^{\prime}$ and YIN yielded large RMS error, corresponding to the strong distortion shown in Fig.~\ref{fig:trajectoryWave}.
The refinement performance without time warping 
is comparable to NDF.
DIO showed the best performance among popular 
methods.
The refined F0 trajectory using F0 adaptive time warping reduces the RMS error by a factor of 10 or more over the range from 2~Hz to 16~Hz modulation.
For higher F0 values, RMS errors of other methods decrease inversely proportionally to the F0 value.

The F0 adaptive time warping also reduced spurious component due to FM substantially.
For example, for a test signal with 16~Hz frequency modulation and 100~musical cent p-p depth,
the refined F0 by the analysis configuration $\mathcal{T}_{10} \circ \mathcal{T}_{10} \circ \mathcal{H}_3$ reduced spurious residual levels lower than $-40$~dB.
This is perceptually negligible.
\ifthenelse{\value{draft}=0}{
}{
{\bf \color{blue} [Remove this entire subsection.]}
 {\bf \color{red}[I removed aperiodicity section and added the last paragraph.]
 }
} 

\section{Discussion}\label{ss:discussion} 
The goal of this paper is to estimate F0 trajectories, which consist of rapidly changing components,  accurately for high-quality speech synthesis.
The proposed set of procedures provide a prospective framework.
However, the following
aspects of F0 estimation were not exploited here.
Investigations of the following issues could be important for improving synthesis quality further.
\ifthenelse{\value{draft}=0}{
}{
{\bf \color{blue} [Start with positive (position?) statement!!]}
{\bf \color{red} [I added the first two sentences and modified the beginning of the third sentence.]}
}

Plosive sounds such as /k/, /t/ sometimes sound like fricative by smearing temporal sharpness due to the smoothing effect of time windowing.
This is a common degradation found in STRAIGHT.

Some speakers and languages frequently use ``creaky voice.'' 
Representing these sounds using periodic signal plus noise results in poor reproduction.
Relevant analysis and representations have to be investigated. 

Temporal variation of F0 consists of effects caused by interactions between harmonic components and group delay in vocal tract transfer function.
It is desirable to compensate this effect for speech synthesis applications, because this effect can be accumulated in each analysis and synthesis cycle.
\ifthenelse{\value{draft}=0}{
}{
 {\bf \color{red} [The following part is a response to your question to the previous version.
 This is not necessary to appear in the final version, I think.]
 
 [When a moving formant trajectory crosses a trajectory of a harmonic component, the group delay around the formant peak modulates the phase of the component. Such modulation are actually visible in F0 trajectories of real speech samples. If this modulated F0 trajectory is used to generate a synthetic sound also using the same formant transition. The same modulation at the crossing of the harmonic component and the formant trajectory occurs and is accumulated.
 
 Worse case is nasal to vowel transition. In the transition, the length of the acoustic tube from glottis to openings (mouth or nostril) changes suddenly and modulates phase of all harmonic component at the same time. This effect is also clearly visible. But, possibly, especially this case, the dips and peaks of F0 trajectory are not audible.
 ]
 }
}

In addition, it is interesting to consider a unique F0 tracker based on {\em Harmonic-Locked Loop} tracking\cite{wang1995ASSP} as an alternative F0 refinement procedure for the third stage of the proposed framework.

  \section{Conclusions}

This paper introduced
a framework
for intantaneous estimates F0 and aperiodicity. 
It is able to improve the ability of F0 extractors to temporally follow varying F0 trajectories by a factor of 10.
It may serve as an useful infrastructure for speech research and applications.
\ifthenelse{\value{draft}=0}{
}{
{\bf \color{dgreen} [In the end, all the sweat about defining aperiodicity in the introduction didn’t seem to matter.  The role of aperiodicity, which it turns out means the energy that is not explained by the dominant harmonic in a particular semitone subband, is as a confidence weight for combining pitch estimates from different bands.  Yes, maybe it could be used for excitation analysis, but we haven’t done that here.

I think this paper should be called “Tracking rapid F0 modulations via weighted subband estimates” or something.  As it stands, you set us up with this general discussion of aperiodicity as if you’re going to solve every problem in speech synthesis, but in fact the scope can be much sharper than that.

But the critical question remains: Do the rapidly-inflected F0s tracked by this technique lead to improved speech copies?  Without the experimental results to confirm what seems like a hunch that this is a significant quality limitation for current synthesis, I think you should downplay the synthesis motivation significantly, and say: here’s a way to track rapid variations in F0.  I’m doing this because I think it’s important to speech synthesis, but for now let’s just focus on this one well-defined problem of recovering the rapid FM I put into the L-F model.]}
{\bf \color{red} [Thank you very much. I will prepare audio samples that will illustrate importance of aperiodicity.]}
}

  \section{Acknowledgements}
  
The authors appreciate insightful discussions with Prof. Roy Patterson on human auditory perception, especially on fine temporal structure and detection of interfering sounds.
Malcolm Slaney provided editorial assistance.
He and Dan Ellis also provided productive as well as critical comments.

\appendix
\ifthenelse{\value{draft}=0}{
\ninept
}{}

\section{Note on the Flanagan's equation}\label{ss:flanagansEq} 
Flanagan uses the time derivative of the logarithm of a complex signal $x(t)$ to estimate the instantaneous frequency.
By introducing a logarithmic function, the phase component is linearly separable from amplitude.
\begin{align}
\!\!\!\!\!\!\!\!\log(x(t)) & \!=\! \log(r(t) \exp(j \theta(t))) 
 = \log(r(t))+j\theta(t) \\
\!\!\!\!\!\Im[\log(x(t))] & = \theta(t) .
\end{align}

To make derivation simpler, as far as no ambiguity is introduced, time dependency representation by $(t)$ is omitted afterwards.
\begin{align}
\omega_i & = \frac{d \theta}{dt} 
 = \Im\left[\frac{d \log(x)}{dt}\right] 
= \Im\left[\frac{1}{x}\frac{dx}{dt}\right] \nonumber \\
& = \Im\left[\frac{\frac{da}{dt}+j\frac{db}{dt}}{a + jb}\right] \ \ \mbox{where} \ \  x = a + jb \nonumber \\
& = \Im\left[\frac{\left(\frac{da}{dt}+j\frac{db}{dt}\right)(a-jb)}
{(a + jb)(a-jb)}\right] \nonumber \\
& = \Im\left[\frac{a\left(\frac{da}{dt}+j\frac{db}{dt}\right)
-jb\left(\frac{da}{dt}+j\frac{db}{dt}\right)}
{a^2+b^2}\right] \nonumber \\
& = \Im\left[\frac{a\frac{da}{dt}+ja\frac{db}{dt}
-jb\frac{da}{dt}-b\frac{db}{dt}}
{a^2+b^2}\right] \nonumber \\
& = \frac{a\dfrac{db}{dt}-b\dfrac{da}{dt}}
{a^2+b^2} 
= \frac{\Re[x]\dfrac{d\Im[x]}{dt}-\Im[x]\dfrac{d\Re[x]}{dt}}{|x|^2} .\label{ew:flanaganIF}
\end{align}
Which is the Flanagan's equation.

The complex-valued signal $x$ in Eq.~\ref{ew:flanaganIF} is a filtered output of $h(t)$.
It is a function of the center frequency $f_c$ and time.
Let explicitly represent $x$ using $X(\omega_c,t)$ and its time derivative using $X_d(\omega_c,t)$.
Then the following holds.
\begin{align}\label{eq:waveletWindows} 
X(t,\omega_c) & = \int_{-\infty}^{\infty}\!\! h(\lambda) x(t-\lambda) d\lambda \nonumber \\
& = -\!\int_{-\infty}^{\infty}\!\!\! w(\tau-t) \exp\left(j\omega_c (\tau-t)\right) x(\tau) d\tau \\
X_d(t,\omega_c) & =
\frac{d X(t, \omega_c)}{dt} \nonumber \\
& =
-\frac{d}{dt}\left(\int_{-\infty}^{\infty}\!\!\! w(\tau-t) \exp\left(j\omega_c (\tau-t)\right) x(\tau) d\tau \right) \nonumber \\
& = -\int_{-\infty}^{\infty} \left(-\frac{d\ w(\tau-t)}{dt} -j\omega_c w(\tau-t)\right) \cdot \nonumber \\
& \ \ \ \ \ \ \ \ \ \ \exp\left(j\omega_c (\tau-t)\right) x(\tau) d\tau \nonumber \\
& = \int_{-\infty}^{\infty} h_d(\lambda) x(t-\lambda) d\lambda ,
\end{align}
where \vspace{-5mm}
\begin{align}
w_d(t) & = \frac{d w(t)}{dt}+j\omega_c w(t) \\
h_d(t) & = w_d(t)\exp(j\omega_c t) .
\end{align}

Substituting these two time windows $w(t)$ and $w_d(t)$ into 
Eq.~\ref{ew:flanaganIF} removes time derivatives:
\begin{align}
\omega_i(t, \omega_c) 
 & = \frac{\Re[X(t,\omega_c)]\Im[X_d(t,\omega_c)]-
 \Im[X(t,\omega_c)]\Re[X_d(t,\omega_c)]}{|X(t,\omega_c)|^2} ,
\end{align}

Note that the TKEO (Teager Kaiser Energy Operator\cite{maragos1993amplitude}) is not relevant for estimating rapidly changing F0 trajectories, since it uses an approximation, which requires slowly changing AM and FM.
Using Flanagan's equation is relevant, since it does not rely on this approximation.

\section{Residual calculation in each detector}\label{ss:residualForFrontend} 
This section shows how the aperiodicity detector in Fig.~\ref{fig:detector} works.
The input to this detector is $x(t)$.
Let $h(t, f_c)$ represent the complex valued impulse response of each band pass filter centered around $f_c$.
\begin{align}
a_k(t,f_c) & = \left|  r(t, f_c) \right|^2 \label{eq:okEq}  \\
r(t, f_c) & = y_1^{\prime}(t, f_c) - y_2^{\prime}(t, f_c) \\
y_2^{\prime}(t, f_c) & = \frac{y_2(t, f_c)}{|y_2(t,f_c)|}  \\
y_2(t, f_c) & = \int_{-2/f_c}^{2/f_c} h(\tau, f_c)y_1^{\prime}(t-\tau) d\tau  \\
y_1^{\prime}(t, f_c) & = \frac{y_1(t, f_c)}{|y_1(t,f_c)|}  \\
y_1(t, f_c) & = \int_{-2/f_c}^{2/f_c} h(\tau, f_c)x(t-\tau) d\tau ,
\end{align}
where the integration interval $(-2/f_c, 2/f_c)$ is for the Nuttall window (Eq.~\ref{eq:bpfIresp}).
For Hann window the interval is $(-1/f_c, 1/f_c)$ and for Blackman window the interval is $(-1.5/f_c, 1.5/f_c)$.
Band pass filters having these impulse response lengths have first spectral zeros at 0 and $2 f_c$.

Smoothing the relative residual level $a_k(t,f_c) $ yields the aperiodicity parameter 
$a_{ks}(t,f_c)$. \vspace{-3mm}
\begin{align}\label{eq:f0DevMes} 
a_{ks}(t,f_c) & =  
 \int_{-2/f_c}^{2/f_c} |h(\tau, f_c)|a_k(t-\tau,f_c) d\tau .
\end{align}

\subsection{Operation and implementation of the procedure}\label{ss:detectorImplementation} 
\begin{figure}
\begin{center}
\includegraphics[width=0.9\hsize]{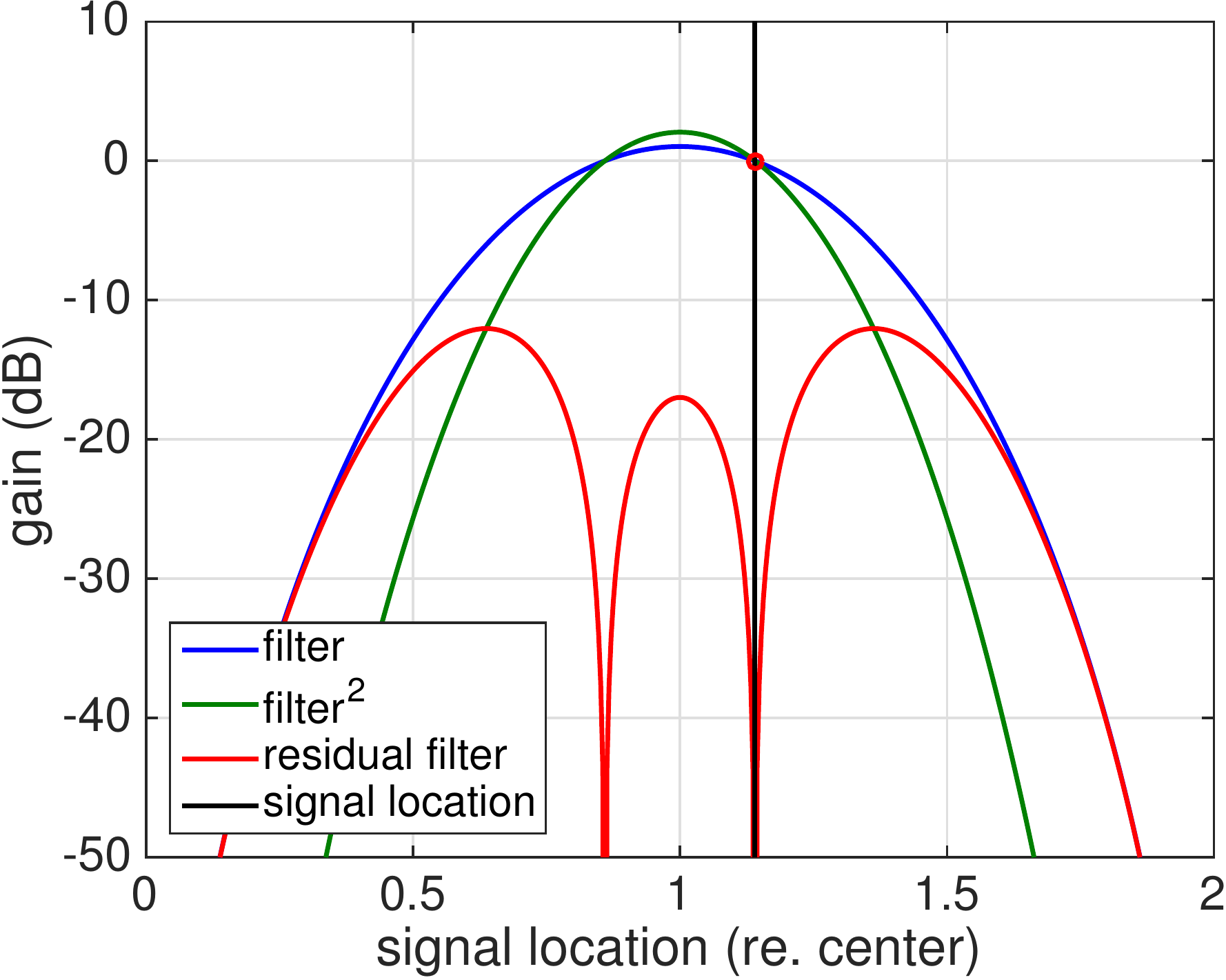} 
\includegraphics[width=0.9\hsize]{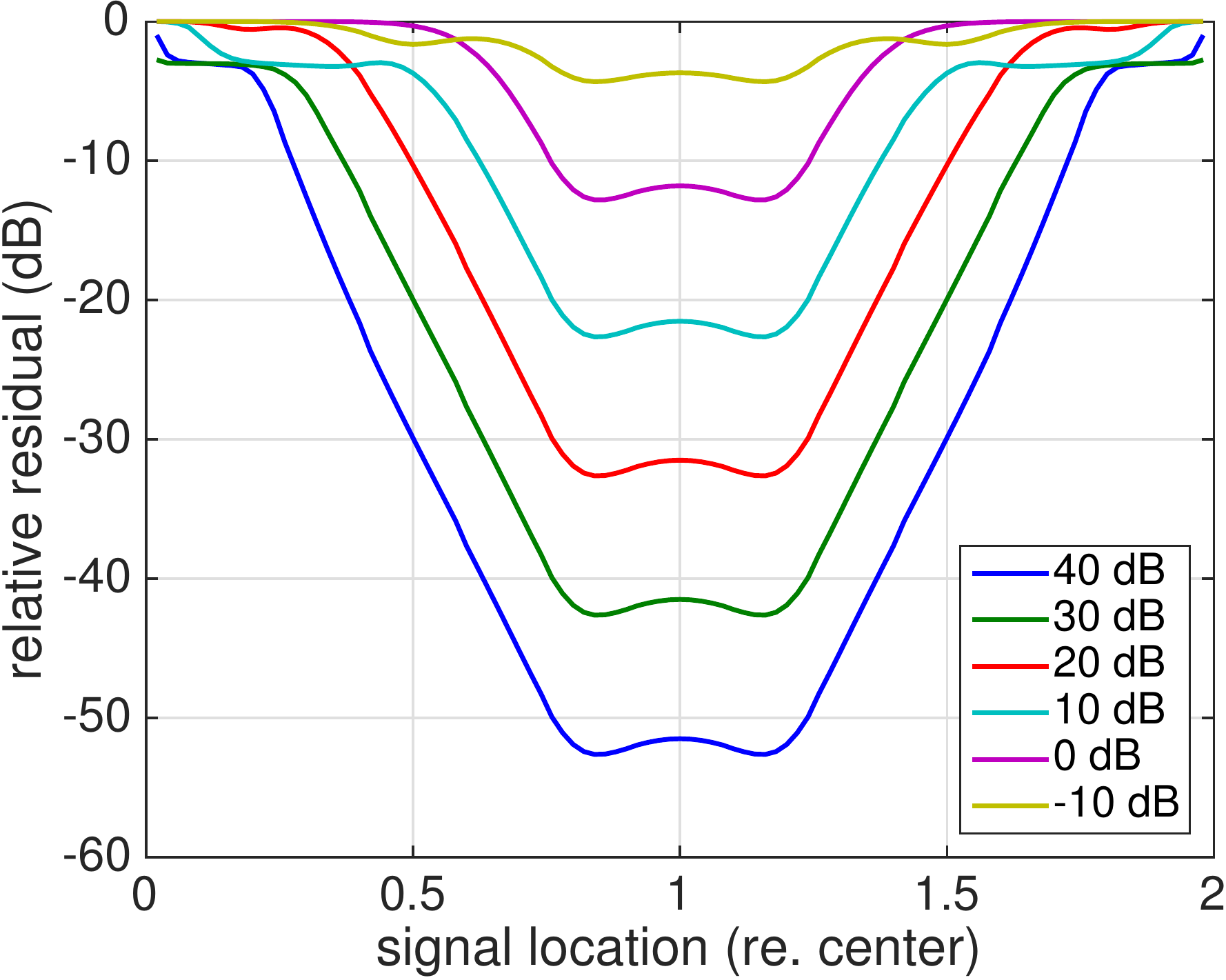} 
\end{center}
\vspace{-6mm}
\caption{principles of operation.
Upper plot shows filter shape and the dominant signal at $1.14 f_c$.
Filter gains are adjusted to make output levels are 0~dB.
Subtracting the second filter gain from the first one yields the
equivalent filter for other components.
Lower plot shows the output residual level as a function of
the location of the dominant signal and the noise level.}
\label{fig:char} 
\ifthenelse{\value{draft}=0}{
\vspace{-4mm}
}{}
\end{figure}
Figure~\ref{fig:char} 
illustrates the process use to calculate the aperiodicity component.
The impulse response of the filter $ h(t, f_c)$ is  \vspace{-3mm}
\begin{align}
w(t) & = \sum_{k=0}^3 a_k \cos(2\pi k f_c t) \ \  |t| < \frac{2}{f_c} \label{eq:bpfIresp}  \\
h(t,f_c) & = w(t) \exp(2\pi j f_c t) , 
\end{align}
where $j=\sqrt{-1}$ and the coefficients $\{a_k\}_{k=0}^3$ are (0.338946, 0.481973, 0.161054, 0.018027). 
This is the 11-th item in Table II of Nuttall's work\cite{nattall1981ieee}.%
\footnote{In terms of time-frequency product, when both is bounded, plorate spheroidal wave function is theoretically the best\cite{slepian1961prolate,slepian1978prolate}.
However, due to large spectral dynamic range of actual speech signals, cosine series windows, which have very low side lobe level and steep side lobe decay\cite{nattall1981ieee} yielded better performance.}

The detector is designed to cancel the primary periodic component in the input signal by adjusting the filter gain at the frequency of the primary component.
This is done by normalizing the output by its RMS level.
In a high SNR case, total RMS level of the filtered signals are approximately equal to the RMS level of the periodic component.
The RMS level of the lower level components are affected by this suppression process.
Since the equivalent filter gain from this suppression process is the difference of two filters, it yields the filter shape shown in the red curve of left plot of Fig.~\ref{fig:char}.
The right plot of Fig.~\ref{fig:char} shows the output aperiodicity parameter $a_k$ as a function of the location of the primary component and the level of the lower level components.

\section{Detector allocation}\label{ss:detectorAllocation} 

\begin{figure}
\begin{center}
\includegraphics[width=0.9\hsize]{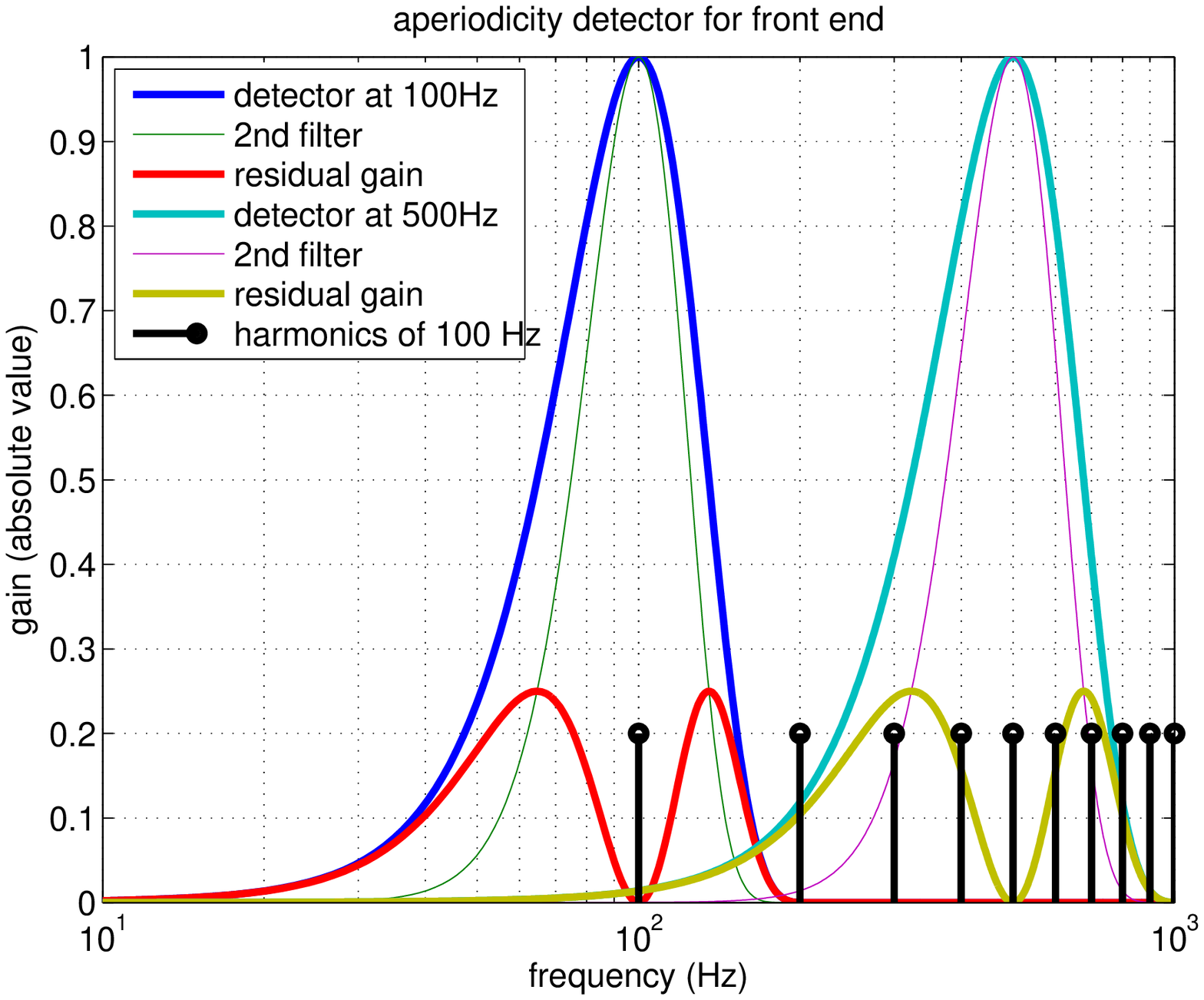} 
\includegraphics[width=0.9\hsize]{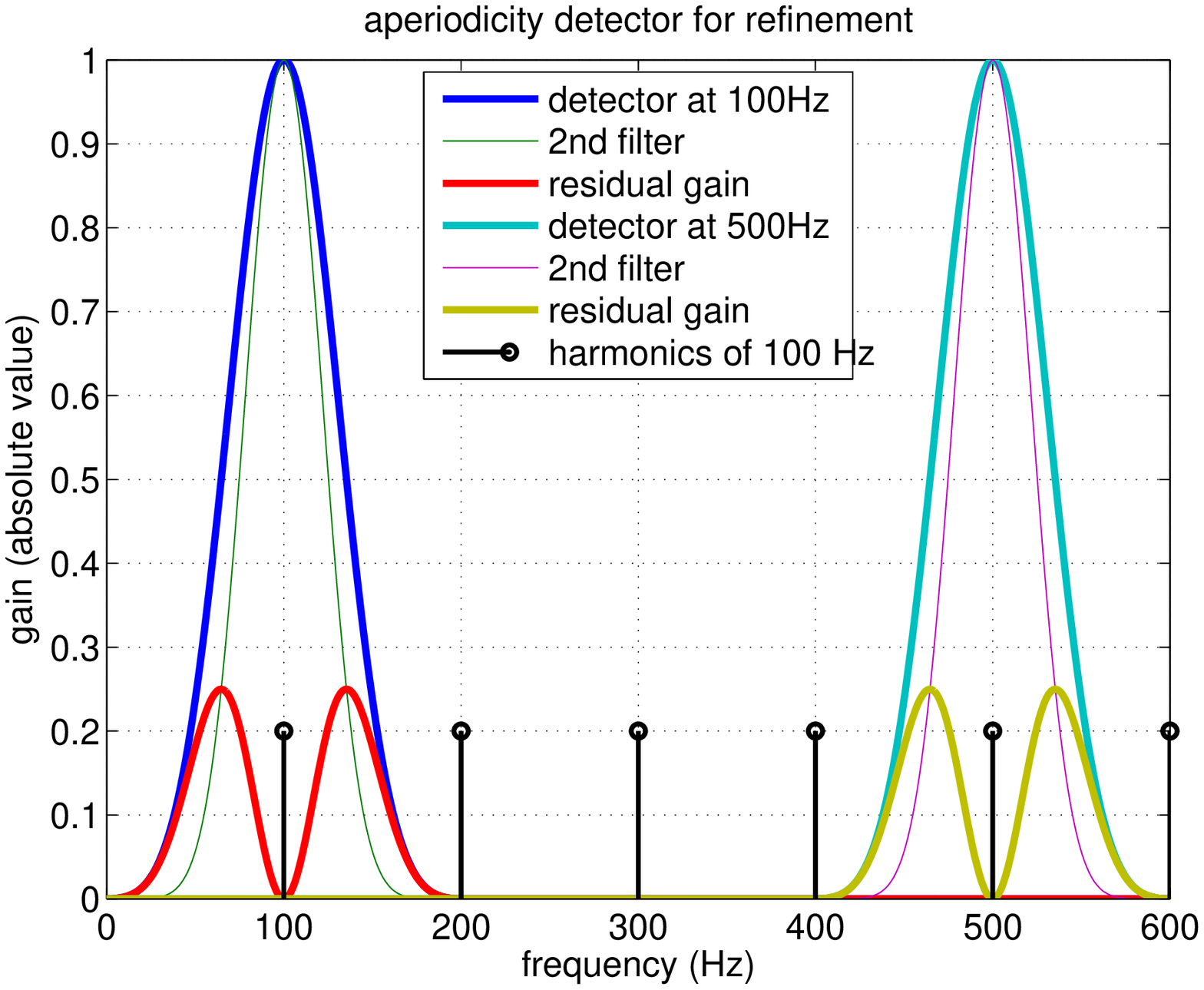} 
\end{center}
\vspace{-6mm}
\caption{Detector allocation of the front end (upper plot) and the refinement stage (lower plot).}
\label{fig:detectorAllocation} 
\ifthenelse{\value{draft}=0}{
\vspace{-4mm}
}{}
\end{figure}

Figure~\ref{fig:detectorAllocation}  shows detector filter shapes of front end and the third stage.
In the front end, the filter width is proportional to the center frequency.
In the refinement stage, the filter width is constant.
The filters in the refinement stage are designed using the estimated F0.

\section{Mixing F0 information}\label{ss:bestMix} 
The band of estimators in the front end independently estimate instantaneous frequency and an estimate of the quality of this estimate in the form of an aperiodicity measure.
We need to consolidate these estimates to get a single estimate of F0 and
we do this with a weighted average.

Assume 
a set of random variables $X_k, k= 1, \ldots, N$ having
zero mean ($\mathbb{E}[X_k]  = 0$) and variances $\sigma_k^2$ ($\mathrm{Var}[X_k] = \sigma_k^2$).
We wish to generate a new estimate from all the noisy estimates by weighting the individual estimates to arrive at an answer with the minimum estimated variance. 
Thus,
assume the following cost function. \vspace{-3mm}
\begin{align}\label{eq:relevantCostM} 
L = \mathrm{Var}\left[\sum_{k=1}^{N} b_k X_k \right] ,
\end{align}
where $b_k$ represents the mixing coefficient.
When mixing F0 estimates derived from different sources, the sum of weights has to satisfy the condition (
$\sum_{k=1}^{N} b_k =  1$).

\ifthenelse{\value{draft}=0}{}
{
******* derivation starts ******\\
This part will not appear in the final version

Assume the following cost function.
\begin{align}\label{eq:relevantCost} 
L = V\left[\sum_{k=1}^{N} b_k X_k \right] ,
\end{align}
where $b_k$ represents the mixing coefficient and has the following constraint
\begin{align}
\sum_{k=1}^{N} b_k = & 1 .
\end{align}
Thus there are $N-1$ free variables, since
\begin{align}\label{eq:nthCoeff} 
b_N = 1-\sum_{k=1}^{N-1}b_k .
\end{align}

Substituting Eq.~\ref{eq:nthCoeff} to Eq.~\ref{eq:relevantCost} yields the following representation of the cost function.
\begin{align}
L = {}
 & V\left[\sum_{k=1}^{N-1} b_k X_k + \left(1-\sum_{k=1}^{N-1}b_k \right) X_N\right] \nonumber \\
 = {}& V\left[ X_N +  \sum_{k=1}^{N-1} b_k (X_k - X_N)\right] \nonumber \\
 = {}&E\left[\left( X_N +  \sum_{k=1}^{N-1} b_k (X_k - X_N)\right)^2\right] \nonumber \\
 = {}   & E\left[X_N^2 + 2 \sum_{k=1}^{N-1} b_k (X_k - X_N)X_N
  + \sum_{k=1}^{N-1} b_k (X_k - X_N) \sum_{m=1}^{N-1} b_m (X_m - X_N)
  \right] \nonumber \\
 = {} & E\left[X_N^2 + 2 \sum_{k=1}^{N-1} b_k X_k X_N - 2 X_N^2 \sum_{k=1}^{N-1} b_k
 + \sum_{k=1}^{N-1}\sum_{m=1}^{N-1} b_m b_k (X_k - X_N)  (X_m - X_N)
\right] \nonumber \\
 = {} & \sigma_N^2 + E\left[
- 2 X_N^2 \sum_{k=1}^{N-1} b_k
 + \sum_{k=1}^{N-1}\sum_{m=1}^{N-1} b_m b_k (X_k - X_N)  (X_m - X_N)
 \right] \nonumber \\
 = {} &\sigma_N^2
 + E\left[- 2 X_N^2 \sum_{k=1}^{N-1} b_k
 + \sum_{k=1}^{N-1}\sum_{m=1}^{N-1} b_m b_k (X_k X_m - X_k X_N - X_m X_N + X_N^2)
 \right] \nonumber \\
 = {}& \sigma_N^2
 - 2 E[X_N^2] \sum_{k=1}^{N-1} b_k
 + \sum_{k=1}^{N-1}\sum_{m=1}^{N-1} b_m b_k (E[X_k X_m] - E[X_k X_N] - E[X_m X_N] + E[X_N^2]) \nonumber \\
 = {}&\sigma_N^2
 - 2 \sigma_N^2 \sum_{k=1}^{N-1} b_k
 + \sum_{k=1}^{N-1}\sum_{m=1}^{N-1} b_m b_k (\delta_{k,m}\sigma_k \sigma_m + \sigma_N^2)
 \nonumber \\
 = {}& \sigma_N^2
 - 2 \sigma_N^2 \sum_{k=1}^{N-1} b_k
 + \sum_{k=1}^{N-1}\sum_{m=1}^{N-1} b_m b_k \sigma_N^2 + \sum_{k=1}^{N-1} b_k^2 \sigma_k^2
\end{align}

At the optimum, all derivatives have to be zero:
\begin{align}
\dfrac{dL}{d b_k} & = \dfrac{d}{db_k}\left[ \sigma_N^2
 - 2 \sigma_N^2 \sum_{p=1}^{N-1} b_p
 + \sum_{p=1}^{N-1}\sum_{m=1}^{N-1} b_m b_p \sigma_N^2 + \sum_{p=1}^{N-1} b_p^2 \sigma_p^2
\right] \nonumber \\
& = - 2 \sigma_N^2 + 2  b_k \sigma_k^2 
+ 2 \sigma_N^2\sum_{p=1}^{N-1} b_p = 0 .
\end{align}

The optimum coefficients $\hat{a}_k$ for $k=1, \ldots, N-1$ are found by solving the following set of equations.
\begin{align}
\sigma_N^2 & =  b_k \sigma_k^2 + \sigma_N^2\sum_{p=1}^{N-1} b_p \ \ \
\ \mbox{for} \ \
(k = 1, \ldots , N-1) .
\end{align}

**** derivation ends ********
}

The optimum coefficients $\hat{b}_k$ for $k=1, \ldots, N-1$ are derived by solving the following set of equations.
\begin{align}\label{eq;bestWeightNm1} 
\sigma_N^2 & =  b_k \sigma_k^2 + \sigma_N^2\sum_{n=1}^{N-1} b_n \ \ \
\ \mbox{for} \ \
(k = 1, \ldots , N-1) .
\end{align}

The final coefficient $\hat{b}_N$ is given by 
\begin{align}\label{eq:bestWeightN} 
\hat{b}_N & = 1 - \sum_{k=1}^{N-1} \hat{b}_k .
\end{align}

Other source of F0 information can be used to improve this estimate further, if the variance of the estimate is available.

\ifthenelse{\value{draft}=0}{
  \eightpt
  }{}
  
  \bibliographystyle{IEEEtran}

  \bibliography{kawaharaIS2016}

\end{document}